\documentclass[aps,prd,10pt,notitlepage,superscriptaddress,nofootinbib,numbers]{revtex4-2}

\usepackage[utf8]{inputenc}
\usepackage[T1]{fontenc}
\usepackage{lmodern}
\usepackage{microtype}
\usepackage{amsmath,amssymb,amsfonts,mathrsfs}
\usepackage{bm}
\usepackage{graphicx}
\usepackage{float}
\usepackage[dvipsnames]{xcolor}
\usepackage{hyperref}
\hypersetup{colorlinks=true,citecolor=Purple,linkcolor=Purple,urlcolor=Purple,breaklinks=true}
\usepackage[capitalize]{cleveref}
\usepackage{tikz}

\definecolor{lime}{HTML}{A6CE39}
\DeclareRobustCommand{\orcidicon}{%
	\begin{tikzpicture}
		\draw[lime, fill=lime] (0,0)
		circle [radius=0.16]
		node[white] {{\fontfamily{qag}\selectfont \tiny ID}};
		\draw[white, fill=white] (-0.0625,0.095)
		circle [radius=0.007];
	\end{tikzpicture}
	\hspace{-2mm}
}

\foreach \x in {A, ..., Z}{%
	\expandafter\xdef\csname orcid\x\endcsname{\noexpand\href{https://orcid.org/\csname orcidauthor\x\endcsname}{\noexpand\orcidicon}}
}

\newcommand\orcidJonathan{{\href{https://orcid.org/0000-0001-9291-0893}{\orcidicon}}}
\newcommand\orcidEdson{{\href{https://orcid.org/0000-0001-9929-5977}{\orcidicon}}}

\usepackage{booktabs}
\usepackage{mathtools}
\newcommand\orcidmca{{\href{https://orcid.org/0009-0009-8703-6092}{\orcidicon}}}
\newcommand\orcidcelio{{\href{https://orcid.org/0000-0002-1266-2218}{\orcidicon}}}
\newcommand{\dd}{\mathrm d}
\newcommand{\EC}{E_{\rm C}}
\newcommand{\PC}{P_{\rm C}}
\newcommand{\Fwh}{\mathcal F}
\newcommand{\Qwh}{\mathcal Q}
\newcommand{\avg}[1]{\left\langle #1\right\rangle}

\begin{document}

\title{Scalar Casimir Effect from Bilocal Correlations in a Wormhole Foam
}

\author{Celio R. Muniz\orcidcelio\!\!}
\email{celio.muniz@uece.br}
\affiliation{Universidade Estadual do Cear\'a (UECE), Faculdade de Educa\c{c}\~ao, Ci\^encias e Letras de Iguatu, Av. D\'ario Rabelo s/n, Iguatu - CE, 63.500-000 - Brasil}

\author{M. C. Ara\'{u}jo\orcidmca}
\email{michelangelo.araujo@ufca.edu.br}
\affiliation{Universidade Federal do Cariri, Av. Tenente Raimundo Rocha, \\
Cidade Universit\'{a}ria, Juazeiro do Norte, Cear\'{a}, 63048-080, Brazil}

\author{Edson Otoniel\orcidEdson\!\!}
\email{edson.otoniel@ufca.edu.br}
\affiliation{Universidade Federal do Cariri (UFCA), Instituto de Forma\c{c}\~ao de Educadores - IFE,  R. Oleg\'ario Emidio de Araujo S/N, Brejo Santo - CE, 63.260-000 - Brazil}

\author{Jonathan A. Rebou\c{c}as\orcidJonathan\!\!}
\email{jalvesreboucas@ifce.edu.br}
\affiliation{Instituto Federal de Educa\c{c}\~ao, Ci\^encia e Tecnologia do Cear\'a (IFCE), Iguatu-CE, Brazil}

\date{\today}

\begin{abstract}
We determine how spatially bilocal correlations induced by a dilute ensemble of virtual Euclidean wormholes modify the scalar Casimir interaction between parallel Dirichlet plates. From the leading monopole action, we isolate an equal-Euclidean-time sector and construct a point-throat, first-order matched spectator model with fixed foam statistics. Global two-plate inclusion-exclusion over all mouth-pair center positions yields a continuous fixed-separation kernel free of the logarithmic image structures produced by a restricted translation sector. The kernel is quadratically suppressed at coincidence, changes sign near 0.695 plate spacings, and tends to a positive large-separation plateau. Averaging over truncated-Gaussian, log-normal, and Gamma distributions with a common relative width yields closely spaced energy curves but a finite shape dependence. Differentiating the dimensional energy at fixed ensemble parameters gives a leading-order pressure threshold of 0.05 in the large-separation limit. In the truncated-Gaussian scan extending to twelve plate spacings, an amplitude of 0.06 yields a crossover near a separation ratio of 3.17, whereas amplitudes of 0.02 and 0.04 remain attractive and 0.05 is critical. The scalar cavity therefore provides a scale-selective probe of spatial bilocality whose leading boundary response can change sign within the declared model.
\end{abstract}

\maketitle

\tableofcontents

\section{Introduction}\label{sec:introduction}

The Casimir effect is among the clearest manifestations of how boundaries alter the spectrum of quantum fields and generate a finite interaction after boundary-independent and isolated-body contributions are removed \cite{Casimir1948,Bordag2001,Milton2001}. Its significance does not require identifying the measured force with an absolute vacuum energy: the observable is the change induced by the boundary configuration, a distinction that is especially important when the Casimir effect is used to test modifications of field propagation or additional weak interactions \cite{Jaffe2005}.

The quantum-field-theoretic treatment of ideal boundaries has clarified both the utility and the limitations of such models. Boundary conditions can be formulated as controlled limits of interactions with background fields, while finite forces between rigid bodies may coexist with nonuniversal surface self-energies \cite{Symanzik1981,GrahamEtAl2002}. Rigorous attraction results also exist for reflection-related configurations under specified positivity assumptions \cite{KennethKlich2006}; consequently, any sign change produced by an additional nonlocal kernel must be interpreted within the assumptions of that effective theory rather than as a contradiction of a universal attraction principle.

Casimir forces have been measured over micrometre and submicrometre ranges with torsion-balance, atomic-force-microscope, and micromachined-oscillator techniques \cite{Lamoreaux1997,MohideenRoy1998,Decca2007}. Relating an ideal calculation to those measurements requires the electromagnetic field and a controlled treatment of optical response, temperature, geometry, roughness, and electrostatic systematics \cite{KlimchitskayaMohideenMostepanenko2009,BezerraKlimchitskayaMostepanenkoRomero2004,Klimchitskaya2021,SedmikPitschmann2021}. In the present work, this experimental record motivates the use of a Casimir cavity as a sensitive boundary observable; it does not provide a direct constraint on the ideal massless-scalar model developed below.

At a more fundamental level, quantum fluctuations of geometry and topology motivated the spacetime-foam picture, beginning with Wheeler's geometrodynamical arguments and later Euclidean path-integral constructions \cite{Wheeler1957,Hawking1978}. Modern assessments emphasize that spacetime foam denotes a family of proposals whose physical content depends on the underlying quantum-gravity framework and whose phenomenological status remains model dependent \cite{Carlip2023}. A gas of virtual wormholes is therefore treated here as a specific effective realization of this broader idea, not as an experimentally established microscopic description of the vacuum.

Euclidean wormholes and baby universes have been studied as mechanisms capable of inducing low-energy interactions and modifying the treatment of couplings or quantum coherence \cite{Hawking1987,Hawking1988Wormholes,GiddingsStrominger1988,GiddingsStrominger1988Axion,Coleman1988,Coleman1988BlackHoles}. This literature also contains longstanding questions about the gravitational path integral, negative modes, factorization, and the Lorentzian interpretation of Euclidean saddles; a contemporary review makes clear that the associated particle-physics and cosmological consequences are conditional on how those issues are resolved \cite{HebeckerMikhailSoler2018}. We accordingly use the Euclidean-wormhole framework as an effective model input and keep its continuation and microscopic interpretation distinct from the boundary calculation itself.

The specific starting point is the statistical program developed by Savelova and by Kirillov and Savelova, in which a dilute ensemble of Euclidean wormholes modifies scalar Green functions and generates a leading bilocal contribution to the scalar effective action \cite{Savelova2015,KirillovSavelova2015}. Subsequent studies in the same program have examined arbitrary wormhole distribution functions, possible fractal or relic populations, and consequences for cosmological, scattering, and gravitational responses \cite{KirillovSavelova2024,KirillovSavelovaVladykina2021,KirillovSavelova2013DarkEnergy,KirillovSavelova2018Scattering,KirillovSavelova2018Newton,KirillovSavelova2025Relic}. These works motivate a statistical treatment of mouth separations, but they do not determine the equal-time reduction, the benchmark separation distributions, or the Casimir kernel adopted in this paper.

Nonlocal quantum field theories require particular care because a Euclidean kernel does not by itself define a causal and unitary Lorentzian theory. Foundational and later analyses show that analyticity, continuation prescriptions, initial-value structure, and the definition of amplitudes are essential to that question \cite{Efimov1967,Tomboulis1997,Tomboulis2015,OsterwalderSchrader1975,BrisceseModesto2019,KoshelevTokareva2021,BrisceseModesto2021}. We therefore restrict the wormhole ensemble to equal Euclidean time so that the retained bilocality is spatial and the effective interaction is local in time after continuation. This is an explicit model reduction: it excludes the complementary unequal-time sector and does not establish reflection positivity or unitarity for the unreduced wormhole theory.

Nonlocality is not new to Casimir physics. Nonlocal interactions concentrated near material boundaries, nonlocal scalar propagators with continuous massive spectra, and boundary effective actions with geometrically organized nonlocal terms have all been investigated \cite{FoscoLosada2009,Saravani2019,BoassoFoscoGuntscheMazzitelli2026}. These studies establish that boundary vacuum energies can respond to nonlocal kernels, but their kernels arise from different physical constructions. To our knowledge, the Casimir response generated specifically by the spatial bilocal kernel of a statistical Euclidean wormhole ensemble, including an average over mouth-separation distributions, has not previously been worked out.

The existing literature connecting Casimir observables and wormholes follows two other directions that must be separated from the present problem. In one, a quantum field is quantized in a prescribed Lorentzian wormhole geometry, so the background spacetime modifies the spectrum or local vacuum energy of the cavity \cite{Khabibullin2006,Santos2020,Sorge2020,MunizBezerraToledo2021}. Here the plates instead remain in an ordinary flat-background boundary problem, while the virtual Euclidean ensemble modifies the scalar two-point structure through correlations between its mouths.

In the other direction, Casimir or Casimir-like negative energies are investigated as sources capable of sustaining Lorentzian wormhole geometries, including long-throat, semiclassical-foam, quantum-gravity-inspired, and modified-gravity constructions \cite{Butcher2014,Garattini2001,Garattini2019,Garattini2020,MunizTangphatiNevesCruz2024,MunizEtAl2025,OliveiraAlencarJardimLandim2022}. The causal role is reversed in our setting: Casimir energy does not support a wormhole, and no individual wormhole is the cavity background. The cavity is used as a probe of an externally specified statistical correlation structure.

Our objective is to determine, within a dilute point-throat and leading-monopole approximation, how this spatially bilocal structure changes the interaction energy and pressure of a massless scalar field between parallel Dirichlet plates. We adopt an explicit matched boundary prescription in which the foam is a fixed spectator ensemble, the scalar Green functions are defined on the plate-separated components, the two-plate inclusion-exclusion difference is formed before the mouth-pair center is integrated over the full normal line, and the result is averaged over a normalized separation density. Three positive-support families matched at the same mean and coefficient of variation are used to measure residual shape dependence. The energy and pressure are kept distinct, and any attraction-repulsion boundary is reported only for the retained leading-order kernel.

The resulting framework provides a scale-selective test of the effective nonlocal correlation rather than a microscopic prediction for the abundance or distribution of wormholes. It also separates what is demonstrated from what remains open: the calculation establishes the scalar Dirichlet response of the equal-time spectator model, whereas an electromagnetic extension, realistic material boundaries, a microscopic derivation of the separation distribution, unequal-time correlations, and foam backreaction are required before contact with experimental bounds or a broader Lorentzian claim can be made.

This paper is organized as follows. Section~II develops the bilocal scalar action, defines the equal-time spatial ensemble, and checks its momentum-space normalization. Section~III constructs the globally subtracted parallel-plate kernel and analyzes its limiting behavior. Section~IV averages that kernel over general and benchmark mouth-separation distributions and obtains the total interaction energy. Section~V derives the physical pressure, assesses perturbative control, identifies the leading-order attraction-repulsion regimes, and discusses the phenomenological scope. Section~VI summarizes the conclusions, while the appendices document the global center integration and the numerical implementation, and compare the matched spectator prescription with an alternative one in which the mouth pairs are confined to the slab. 

\section{Bilocal scalar action and equal-time spatial reduction}
\label{sec:effective}

In this section we begin with the free scalar field and then introduce the leading
bilocal monopole inherited from the Euclidean-foam construction.  Only after
that source term is stated do we define the equal-time isotropic ensemble,
derive its momentum-space normalization, and formulate the first-order vacuum
correction.  Boundary assumptions are postponed to the next section.
Throughout this section we use Euclidean signature and natural units,
$\hbar=c=1$, and write $\mathsf L$ for the physical dimension of length.

\subsection{Bilocal effective action and equal-time reduction}
\label{sec:bilocal_reduction}

We consider a massless real scalar field $\phi(Y)$ in four-dimensional
Euclidean space, where $Y$ is a Euclidean spacetime point and
$M\in\{1,2,3,4\}$ is a Euclidean index, with free action
\begin{equation}
 S_0[\phi]
 =
 -\frac12\int \dd^4Y\,
 \phi\partial_M^2\phi
 =
 \frac12(\phi,K_0\phi),
 \qquad
 K_0=-\Delta_E,
\label{eq:S0}
\end{equation}
and $\Delta_E = \partial_M^2$. For two Euclidean points $Y$ and $Y'$, the corresponding free Green function
satisfies
\begin{equation}
 -\Delta_E G_0(Y-Y')
 =
 \delta^{(4)}(Y-Y').
 \label{eq:G0eq}
\end{equation}
For the nonzero displacement $\Delta Y\equiv Y-Y'$, its four-dimensional kernel is
\begin{equation}
 G_0(\Delta Y)=\frac{1}{4\pi^2(\Delta Y)^2}.
 \label{eq:G0}
\end{equation}
Equation~\eqref{eq:G0} is the kernel for $\Delta Y\neq0$; its coincidence
limit is singular.  Equation~\eqref{eq:G0eq} holds in the distributional
sense, and only finite subtracted combinations are retained when the vacuum
observable is constructed.

Now, let $a>0$ be the throat radius, $R_\pm$ the two mouth positions,
$\xi$ the collection of wormhole parameters, and $\rho(\xi)\geq0$ the
ensemble density with measure $\dd\xi$.
In the dilute-wormhole approximation, the leading monopole
contribution to the scalar effective action derived by Kirillov and
Savelova is bilocal \cite{KirillovSavelova2015},
\begin{equation}
 V_1[\phi]
 =
 \pi^2\int \dd\xi\,a^2
 \left[\phi(R_+)-\phi(R_-)\right]^2
 \rho(\xi).
 \label{eq:V1original}
\end{equation}
In the full geometric construction a wormhole may also
carry a relative gluing orientation between its mouths. At the
lowest-monopole level retained in Eq.~\eqref{eq:V1original}, however,
the scalar contribution is independent of this orientation, so that
the relevant parameters may be represented by
$\xi=(a,R_+,R_-)$ \cite{Savelova2015,KirillovSavelova2015}.
The essential feature of Eq.~\eqref{eq:V1original} is therefore that
the foam does not introduce an additional local scalar degree of
freedom: it couples two values of the same field evaluated at the
wormhole mouths.

For a homogeneous Euclidean ensemble, specialize $Y$ to the position of a
reference mouth and define the mouth-separation four-vector
$X\equiv R_+-R_-$.
Translational invariance then gives
\begin{equation}
 V_1[\phi]
 =
 \pi^2\int
 a^2
 \left[\phi(Y+X)-\phi(Y)\right]^2
 \rho(a,X)\,
 \dd a\,\dd^4X\,\dd^4Y .
 \label{eq:V1homogeneous}
\end{equation}
Since the scalar field has $[\phi]=\mathsf L^{-1}$, dimensional
consistency requires the four-dimensional ensemble density to have
$[\rho]=\mathsf L^{-9}$.  With the measure
$\dd a\,\dd^4X$, the integrated event intensity consequently has
$[n_4]=\mathsf L^{-4}$, where
\begin{equation}
 \ n_4\equiv\int \rho(a,X)\,\dd a\,\dd^4X. 
 \label{eq:n4}
\end{equation}

Our purpose is to isolate a sector in which the foam-induced
interaction remains spatially nonlocal but becomes local in time
after Wick rotation. We therefore do not identify the original
$O(4)$ ensemble with an $O(3)$ one (O(n) is the n-dimensional orthogonal group); instead, we write
$X=(X_4,\mathbf r)$ and define the reduced equal-Euclidean-time ensemble by
imposing
\begin{equation}
 \rho_{\rm ET}(a,X)
 =
 \delta(X_4)\,
 \widehat\rho(a,\mathbf r).
 \label{eq:rhoETdelta}
\end{equation}
Because $[\delta(X_4)]=\mathsf L^{-1}$, one has
$[\widehat\rho]=\mathsf L^{-8}$.  
Thus, the reduced event intensity, defined as
\begin{equation}
 n_{\rm ET}
 \equiv
 \int
 \widehat\rho(a,\mathbf r)\,
 \dd a\,\dd^3r,
 \label{eq:nETdef}
\end{equation} 
has $[n_{\rm ET}]=\mathsf L^{-4}$.  
Therefore, $n_{\rm ET}$ is not a three-dimensional number density of
persistent wormholes. It is the intensity of the delta-supported
Euclidean ensemble selected by Eq.~\eqref{eq:rhoETdelta}, and in
general it need not be numerically equal to $n_4$.

For identical throats of radius $a$, retained at leading monopole order,
and an isotropic distribution of the spatial mouth separation, let $P(r)\geq0$ denote
the normalized radial probability density, \begin{equation}
 \int_0^\infty P(r)\,\dd r=1,
 \label{eq:Pnorm}
\end{equation}
and let $a'>0$ be the throat-radius
integration variable. Thus, we write
\begin{equation}
 \widehat\rho(a',\mathbf r)
 =
 n_{\rm ET}\,
 \delta(a'-a)\,
 \frac{P(r)}{4\pi r^2},
 \qquad
 r=|\mathbf r|,
 \label{eq:rhoETfactor}
\end{equation}
This describes the separation of the two mouths within the reduced ensemble,
rather than a microscopic prediction for the foam distribution. More
general wormhole distributions have been considered in the Euclidean
foam framework \cite{KirillovSavelova2024}; here $P(r)$
remains arbitrary. It is important to emphasize that finite spherical throats can be nonoverlapping only if $r\geq2a$. Here, we use point-throat geometry in the radial average, extend the integral to
$r=0$, and retain only the leading $a^2$ monopole coupling.  For finite $a$,
the excluded probability $\int_0^{2a}P(r)\,\dd r$ is not quantified by the
present reduction.

Also, it is useful to let $\tau$ denote Euclidean time and to choose the mouth
midpoint $\mathbf R$ as the spatial center coordinate,
\begin{equation}
 R_\pm
 =
 \left(
 \tau,\,
 \mathbf R\pm\frac{\mathbf r}{2}
 \right).
 \label{eq:mouthcoords}
\end{equation}
Let $\dd\Omega_{\hat r}$ be the solid-angle measure of the unit vector
$\hat r=\mathbf r/r$; division by $4\pi$ implements isotropy.
Substituting Eqs.~\eqref{eq:rhoETdelta} and
\eqref{eq:rhoETfactor} into Eq.~\eqref{eq:V1homogeneous} then gives
\begin{equation}
 V_1[\phi]=\pi^2n_{\rm ET}a^2\int\dd\tau\,\dd^3R\int_0^\infty\dd r\,P(r)\int\frac{\dd\Omega_{\hat r}}{4\pi}\left[\phi\left(\tau,\mathbf R+\frac{\mathbf r}{2}\right)-\phi\left(\tau,\mathbf R-\frac{\mathbf r}{2}\right)\right]^2 .
 \label{eq:V1reduced}
\end{equation}
From this point onward we abbreviate $n\equiv n_{\rm ET}\geq0$. Notice
that the combination $\dd\tau\,\dd^3R\,n a^2\phi^2$ is dimensionless, providing an
immediate dimensional check of the reduced action. Also note that the restriction in Eq.~\eqref{eq:rhoETdelta} is therefore a
modeling choice rather than a claim that the original Euclidean
wormhole ensemble is intrinsically $O(3)$. It selects the
equal-time spatially nonlocal sector retained in
the effective theory.

\subsection{Momentum-space representation and normalization check}
\label{sec:deltaKderivation}

The momentum-space representation of Eq.~\eqref{eq:V1reduced} provides
an independent normalization check and identifies the scale dependence of
the quadratic insertion. Let $k_4\in\mathbb R$ be the Euclidean frequency and
$k\equiv|\mathbf k|\geq0$ the spatial wave-number magnitude.
We use the Euclidean Fourier convention
\begin{equation}
 \phi(\tau,\mathbf R)
 =
 \int
 \frac{\dd k_4\,\dd^3k}{(2\pi)^4}
 \widetilde\phi(k_4,\mathbf k)
 e^{ik_4\tau+i\mathbf k\cdot\mathbf R}.
 \label{eq:Fourierphi}
\end{equation}
The field difference entering the bilocal interaction becomes
\begin{equation}
 \phi\left(\tau,\mathbf R+\frac{\mathbf r}{2}\right)-\phi\left(\tau,\mathbf R-\frac{\mathbf r}{2}\right)=2i\int\frac{\dd k_4\,\dd^3k}{(2\pi)^4}\widetilde\phi(k_4,\mathbf k)e^{ik_4\tau+i\mathbf k\cdot\mathbf R}\sin\left(\frac{\mathbf k\cdot\mathbf r}{2}\right).
 \label{eq:mouthdiffFourier}
\end{equation}
After squaring this expression and integrating over $\tau$ and
$\mathbf R$, the momentum-conserving delta functions collapse the
second Fourier integral. Using
$4\sin^2(\mathbf k\cdot\mathbf r/2)=
2[1-\cos(\mathbf k\cdot\mathbf r)]$ and comparing the result with the
standard quadratic form, one obtains
\begin{equation}
 V_1
 =
 \frac12
 \int
 \frac{\dd k_4\,\dd^3k}{(2\pi)^4}
 \widetilde\phi(-k_4,-\mathbf k)\,
 \delta K(\mathbf k)\,
 \widetilde\phi(k_4,\mathbf k),
 \label{eq:deltaVquadratic}
\end{equation}
where
\begin{equation}
 \delta K(\mathbf k)
 =
 4\pi^2na^2
 \int_0^\infty\dd r\,P(r)
 \int\frac{\dd\Omega_{\hat r}}{4\pi}
 \left[
 1-\cos(\mathbf k\cdot\mathbf r)
 \right].
 \label{eq:deltaKbeforeangle}
\end{equation}
For an isotropic distribution, the angular average is the spherical
Bessel function of order zero $j_0$,
\begin{equation}
 \int
 \frac{\dd\Omega_{\hat r}}{4\pi}
 e^{i\mathbf k\cdot\mathbf r}
 =
 j_0(kr)
 =
 \frac{\sin(kr)}{kr}.
 \label{eq:sphericalBessel}
\end{equation}
Now, defining the isotropic spatial form factor of the mouth-separation distribution
by
\begin{equation}
 \mathcal S(k)
 =
 \int_0^\infty
 \dd r\,P(r)
 \frac{\sin(kr)}{kr},
 \label{eq:structurefactor}
\end{equation}
we can write the quadratic insertion as
\begin{equation}
 \delta K(\mathbf k)
 =
 4\pi^2na^2
 \left[
 1-\mathcal S(k)
 \right].
 \label{eq:deltaK}
\end{equation}
This insertion therefore has $[\delta K]=\mathsf L^{-2}$, matching the
dimension of the free quadratic kernel in Eq.~\eqref{eq:G0eq}.

Equation~\eqref{eq:deltaK} also provides a direct normalization check
against the original four-dimensional foam result. For a homogeneous
wormhole distribution, let
$\mathfrak k=(k_4,\mathbf k)$ denote Euclidean four-momentum and let
$\rho(a',\mathfrak k)$ denote the Fourier transform of the separation
density. Kirillov and Savelova write the leading
momentum-space correction in the form \cite{KirillovSavelova2015}
\begin{equation}
 \nu(\mathfrak k)
 =
 4\pi^2
 \int\dd a'\,a'^2
 \left[
 \rho(a',0)-\rho(a',\mathfrak k)
 \right].
 \label{eq:nuKS}
\end{equation}
Applying the equal-time distribution of
Eqs.~\eqref{eq:rhoETdelta}-\eqref{eq:rhoETfactor} gives
\begin{equation}
 \nu_{\rm ET}(\mathfrak k)
 =
 4\pi^2na^2
 \left[
 1-\mathcal S(k)
 \right]
 =
 \delta K(\mathbf k).
 \label{eq:nuETcheck}
\end{equation}
The right-hand side is independent of $k_4$ because the imposed
equal-time density is supported at $X_4=0$.
Thus the equal-time reduction introduces no additional numerical
factor into the leading quadratic insertion. This agreement also
provides an independent check of the normalization of
Eq.~\eqref{eq:V1reduced}.
The calculation retains the full spatial form factor
$1-\mathcal S(k)$ rather than a finite derivative expansion.  It therefore
does not identify the kernel with a Pauli-Villars regulator or draw a
Lorentzian stability conclusion from the Euclidean expression.

The insertion $\delta K(\mathbf k)$ characterizes the translationally
invariant reduced ensemble in unbounded space.  We now express the same
first-order term through a background-generic bilocal Green-function
combination.

\subsection{First-order vacuum functional}
\label{sec:firstorder_vacuum}

Let
\begin{equation}
 \mathcal Z
 =
 \int\mathcal D\phi\,
 e^{-S_0[\phi]-V_1[\phi]},
 \qquad
 W=-\ln \mathcal Z ,
 \label{eq:vacuumfunctional}
\end{equation}
$W_0$ denote $W$ evaluated at $V_1=0$, and $\langle\cdot\rangle_0$ the expectation in that Gaussian theory.
Expanding to first order in the dilute-foam interaction gives
\begin{equation}
 W
 =
 W_0+\avg{V_1}_0+O(V_1^2)
 \qquad \implies \qquad
 \Delta W^{(1)}
 =
 \avg{V_1}_0.
 \label{eq:firstorderW}
\end{equation}
  At this stage the Green function $G$ is left
unspecified; the boundary geometry is introduced in the next section.

For a fixed pair of mouths,
\begin{equation}
\begin{aligned}
 \avg{[\phi(R_+)-\phi(R_-)]^2}_0
 &=\avg{\phi(R_+)^2}_0+\avg{\phi(R_-)^2}_0-2\avg{\phi(R_+)\phi(R_-)}_0\\
 &=G(R_+,R_+)+G(R_-,R_-)-2G(R_+,R_-).
\end{aligned}
 \label{eq:bilocalG}
\end{equation}
Hence, at first order in the event intensity and at leading
monopole order in $a^2$, the dependence on a subsequently chosen
background is encoded in the same bilocal Green-function combination,
\begin{equation}
 D_G(R_+,R_-)
 \equiv
 G(R_+,R_+)
 +
 G(R_-,R_-)
 -
 2G(R_+,R_-).
 \label{eq:DGdef}
\end{equation}
Substitution into Eq.~\eqref{eq:V1reduced} gives
\begin{equation}
 \Delta W^{(1)}=\pi^2na^2\int\dd\tau\,\dd^3R\int_0^\infty\dd r\,P(r)\int\frac{\dd\Omega_{\hat r}}{4\pi}D_G\left(R_+,R_-\right).
 \label{eq:DeltaW1}
\end{equation}

For a static configuration of Euclidean duration $T_E$, the associated
vacuum-energy correction is
\begin{equation}
 \Delta E^{(1)}
 =
 \lim_{T_E\to\infty}
 \frac{\Delta W^{(1)}}{T_E}.
 \label{eq:DeltaEfromW}
\end{equation}
It is important to highlight that equation~\eqref{eq:DeltaW1} is background-generic and is not yet a
Casimir observable.  The next section specifies the boundary Green
functions and defines the corresponding finite interaction subtraction.

\section{Parallel-plate cavity and the fixed-separation kernel}
\label{sec:cavity}

Section~\ref{sec:effective} fixed the free field, the equal-time
ensemble, and its leading bilocal interaction without boundaries.  We now
add the boundary-matching postulate: $S_0+V_1$ is used as a quadratic
effective action for a scalar field obeying Dirichlet conditions, the
wormhole distribution is held fixed, the scalar Gaussian average is taken,
and the two-plate inclusion-exclusion difference is formed last.  We call
this the matched spectator model.  

\subsection{Dirichlet Green function and interaction subtraction}
\label{sec:dirichlet_subtraction}

Let $z$ be the Cartesian coordinate normal to the plates, $L>0$ their
separation, and $A$ their area, with the infinite-area limit understood.
The ideal Dirichlet plates occupy $z=0$ and $z=L$, where
$\phi|_{z=0}=\phi|_{z=L}=0$.
Energies and forces will be reported per unit area.
For reference, the standard massless scalar Casimir energy and
pressure are \cite{Bordag2001,Milton2001}
\begin{equation}
 \frac{\EC^{(0)}}{A}
 =
 -\frac{\pi^2}{1440L^3},
 \qquad
 \PC^{(0)}
 =
 -\frac{\pi^2}{480L^4}.
 \label{eq:ordinaryCasimir}
\end{equation}

For Euclidean points
$\mathsf x=(\tau,\mathbf x_\parallel,z)$ and
$\mathsf x'=(\tau',\mathbf x'_\parallel,z')$, write
$\Delta\tau=\tau-\tau'$ and
$\Delta\mathbf x_\parallel=\mathbf x_\parallel-\mathbf x'_\parallel$.
The Euclidean Dirichlet Green
function between the two plates can be represented by the image
series, where $m\in\mathbb Z$ labels the images,
\begin{equation}
 G_{12}(\mathsf x,\mathsf x')=\sum_{m=-\infty}^{\infty}\Big[G_0(\Delta\tau,\Delta\mathbf x_\parallel,z-z'+2mL)-G_0(\Delta\tau,\Delta\mathbf x_\parallel,z+z'+2mL)\Big].
 \label{eq:GDimages}
\end{equation}
The first term generates the translational images, whereas the second
implements the reflections required by the Dirichlet boundaries. The Casimir interaction must be isolated from the free-space vacuum
contribution and from the self-energies associated with either plate
separately.  Let $G_1$ and $G_2$ denote the Green functions for the
isolated plates at $z=0$ and $z=L$, respectively.
We therefore introduce the inclusion-exclusion
combination
\begin{equation}
 G_{\rm int}
 \equiv
 G_{12}-G_1-G_2+G_0.
 \label{eq:Gintdef}
\end{equation}
In particular, the ultraviolet-divergent local coincidence
piece is common to the four Green functions and cancels from
Eq.~\eqref{eq:Gintdef}.

For the bilocal interaction derived in Sec.~\ref{sec:effective}, the
relevant quantity is not a single Green function but the combination
in Eq.~\eqref{eq:DGdef}. The boundary-dependent counterpart is
therefore
\begin{equation}
 \mathcal D_{\rm int}(R_+,R_-)=G_{\rm int}(R_+,R_+)+G_{\rm int}(R_-,R_-)-2G_{\rm int}(R_+,R_-).
 \label{eq:Dint}
\end{equation}
The inclusion-exclusion subtraction must be formed for this complete
bilocal combination before any image or spatial sector is discarded.
This ordering is essential because the cross-mouth term
$G_{\rm int}(R_+,R_-)$ probes configurations in which the two mouths
can occupy different regions separated by the plates.

The Green functions in Eq.~\eqref{eq:Gintdef} are defined component by
component.  Equation~\eqref{eq:GDimages} applies inside the slab; exterior
components and the one-plate problems use the corresponding single-image
Green functions, while propagation across an ideal Dirichlet plate is set
to zero.  This prescription specifies the boundary model, but it is an
imposed extension of the homogeneous bilocal kernel: the source construction
does not determine whether a virtual mouth pair couples scalar regions that
an ideal plate disconnects.  An alternative prescription, in which only
pairs with both mouths inside the slab are retained, is compared in
Appendix~\ref{app:confined}; it preserves the sign structure of the kernel but
changes the quantitative thresholds.

For a nonnegative normalized $P(r)$, Eq.~\eqref{eq:deltaK} is
nonnegative in unbounded Euclidean space.  This does not determine the sign
of the Casimir interaction because the inclusion-exclusion operation in
Eq.~\eqref{eq:Gintdef} acts on a difference of Green functions and need not
preserve the sign of the free insertion.

Under the boundary-matching prescription stated
at the beginning of this section, the plates constrain the scalar field but
do not confine or rearrange the virtual-wormhole ensemble. The center
of a mouth pair is consequently integrated over the full normal coordinate.
Let $z_c$ denote the normal component of that center, with domain
$-\infty<z_c<\infty$,
rather than only over $0<z_c<L$. Configurations with two mouths
inside the slab, one mouth inside and one outside, and both mouths
outside must therefore all be retained. Their explicit geometric
decomposition is given in Appendix~\ref{app:global}.

\subsection{Geometry of a spatial mouth pair}
\label{sec:pair_geometry}

For a fixed mouth separation $r$, let $\theta$ denote the angle
between $\mathbf r$ and the normal to the plates. We define
\begin{equation}
 \mu=\cos\theta,
 \qquad
 \frac{\dd\Omega_{\hat r}}{4\pi}
 =
 \frac12\,\dd\mu .
 \label{eq:S2measure}
\end{equation}
Reflection symmetry about the plate normal allows the angular average
to be restricted to $0\leq\mu\leq1$. It is convenient to introduce
the dimensionless variables
\begin{equation}
 x\equiv\frac{r}{L},
 \qquad
 q\equiv x\mu,
 \qquad
 p\equiv x\sqrt{1-\mu^2},
 \label{eq:xpq}
\end{equation}
where $q$ and $p$ are respectively the normal and parallel
mouth separations in units of $L$.
The integration domain is $x\geq0$, $0\leq\mu\leq1$,
$0\leq q\leq x$, and $0\leq p\leq x$.  The endpoint $x=0$ is the
formal coincident-mouth limit of the point-throat model; it is not allowed
for two finite nonoverlapping throats.
Because $0\leq\mu\leq1$, Eq.~\eqref{eq:xpq} gives a nonnegative normal
separation $q$.

The dimensionless center coordinate and the two normal mouth coordinates are
\begin{equation}
 Z\equiv\frac{z_c}{L},
 \qquad
 \frac{z_\pm}{L}
 =
 Z\pm\frac{q}{2}.
 \label{eq:zpm}
\end{equation}
This makes the origin of the two geometric sectors transparent. For
$q<1$, both mouths lie between the plates when
$q/2<Z<1-q/2$. At $q=1$ this interval
shrinks to zero, while for $q>1$ the two mouths can instead straddle
the complete slab, with one below $z=0$ and the other above $z=L$.
The global center integral must therefore be evaluated separately for
$q<1$ and $q\geq1$. The complete interval decomposition, including
the one-inside/one-outside and both-outside sectors, is displayed in
Appendix~\ref{app:global}.

\subsection{Global fixed-separation kernel}
\label{sec:fixed_kernel}

For a monodisperse mouth separation $r$, Eq.~\eqref{eq:DeltaEfromW}
and the interaction subtraction above give, to first order in the
event intensity and at leading order in $a^2$, an energy per unit area that
we express through the dimensionless fixed-pair response
$\Fwh_{\rm glob}(x)$ as
\begin{equation}
 \frac{\Delta\EC(r)}{A}
 =
 \frac{na^2}{L}\,
 \Fwh_{\rm glob}(x),
 \qquad
 x=\frac{r}{L}.
 \label{eq:mainfixed}
\end{equation}
The overall scaling follows directly from dimensional analysis:
$[na^2/L]=\mathsf L^{-3}$, as required for an energy per unit area.

%
From here on, distances are measured in units of $L$.
The free propagator is $G_0=1/(4\pi^2d^2)$, where $d$ is the distance between
the two points.  To remove the factor $4\pi^2$ and the dimension of $G$, we
work with the dimensionless combination
\begin{equation*}
 S_{ab}\equiv4\pi^2L^2\,G_{\rm int}(R_a,R_b),
 \qquad a,b\in\{+,-\}.
\end{equation*}
In this combination, each free-propagator term of $G_{\rm int}$ (between a
mouth and one of its images) contributes simply $1/\ell^2$, where $\ell=d/L$ is
the distance in units of $L$; for example, a distance $d=2L$ contributes
$1/4$.  We insert Eq.~\eqref{eq:Dint} into
Eq.~\eqref{eq:DeltaW1} and divide by $T_E$ as in
Eq.~\eqref{eq:DeltaEfromW}.  The integral over the directions parallel to
the plates gives the area $A$, the angular average equals $\int_0^1\dd\mu$
by reflection symmetry (Eq.~\eqref{eq:S2measure}), and we write $z_c=LZ$.
Comparison with Eq.~\eqref{eq:mainfixed} then gives, for fixed $r$,
\begin{equation*}
 \Fwh_{\rm glob}(x)
 =\frac14\int_0^1\dd\mu\int_{-\infty}^{\infty}\dd Z\,
 \big[S_{++}+S_{--}-2S_{+-}\big].
\end{equation*}
Two kinds of terms appear.  The coincident-mouth terms do not depend on
$r$: their global integral is
$\int\dd Z\,S_{\pm\pm}=\pi^2/12$ for each mouth (the slab contributes
$\pi^2/12-\tfrac12$ and the two exterior regions $\tfrac12$ together; see
Appendix~\ref{app:global}).
The cross-mouth term defines the function
\begin{equation*}
 C(x,\mu)\equiv\int_{-\infty}^{\infty}\dd Z\,S_{+-}(Z;p,q),
\end{equation*}
the globally integrated correlation between the two mouths.
The functions defined next are the pieces of $C$ obtained by splitting the
$Z$ integral according to where the two mouths lie relative to the plates
(see Appendix \ref{app:global}): $T$ is the sum over translational images
when both mouths are inside the slab, minus the free-space term;
$\mathcal R_{\rm refl}$ is the sum over reflections in the two plates, minus
the single-plate reflections already removed by
Eq.~\eqref{eq:Gintdef}; and the arctangent terms come from the sectors in
which at least one mouth lies outside the slab.

For $q<1$, the translational-image contribution is
\begin{equation}
 T(p,q)
 =
 \frac{\pi}{2p}
 \frac{\sinh(\pi p)}
 {\cosh(\pi p)-\cos(\pi q)}
 -\frac{1}{x^2},
\label{eq:Tpq}
\end{equation}
and the reflected residual is
\begin{equation}
 \mathcal R_{\rm refl}(p,q)=\int_{q/2}^{1-q/2}\dd Z\left[\frac{\pi}{2p}\frac{\sinh(\pi p)}{\cosh(\pi p)-\cos(2\pi Z)}-\frac{1}{p^2+4Z^2}-\frac{1}{p^2+(2Z-2)^2}\right].
\label{eq:Rpq}
\end{equation}
The global center integration then yields
\begin{equation}
 C_{<}(x,\mu)
 =
 \frac{1}{p}
 \tan^{-1}\left(\frac{p}{2-q}\right)
 +(1-q)T(p,q)-\mathcal R_{\rm refl}(p,q).
 \label{eq:Cless}
\end{equation}
The factor $(1-q)T$ arises from the translational-image contribution
over the interval in which both mouths lie inside the slab, whereas
$\mathcal R_{\rm refl}$ is the reflected residual left by the complete two-plate
subtraction. The remaining arctangent term collects the sectors with at
least one mouth outside the slab: for $q<1$ these are the four exterior
sectors of Table~\ref{tab:center_sectors}, in which $S_{+-}$ reduces to a
single image term, and no straddling sector exists in this branch. Appendix
\ref{app:global} gives the corresponding interval-by-interval
construction.

For $q\geq1$, no both-inside interval remains and the global center
integration reduces to
\begin{equation}
 C_{>}(x,\mu)
 =
 \frac{\arccos\mu}
 {x\sqrt{1-\mu^2}}
 +\frac{\mu}{x}
 -\frac{1}{x^2}.
 \label{eq:Cgreater}
\end{equation}
The apparent factors $1/p$ and $1/\sqrt{1-\mu^2}$ for an orientation
parallel to the plate normal are evaluated through their continuous limits
as $p\to0$ (equivalently $\mu\to1$); the combined branch expressions are
finite.
The two branches join continuously at $q=1$. Indeed,
\begin{equation}
 C_{<}\big|_{q=1}
 =
 \frac{\arctan\sqrt{x^2-1}}
 {\sqrt{x^2-1}}
 =
 \frac{\arccos(1/x)}
 {\sqrt{x^2-1}}
 =
 C_{>}\big|_{q=1}.
 \label{eq:Cmatching}
\end{equation}
This establishes equality of the branch values at the sector boundary; it
does not by itself prove differentiability of the angular integrand.

With both branches now defined, introduce the composite angular kernel
\begin{equation}
 C(x,\mu)
 =
 \begin{cases}
 C_{<}(x,\mu), & q<1,\\[2mm]
 C_{>}(x,\mu), & q\geq1.
 \end{cases}.
 \label{eq:Cpiecewise}
\end{equation}
For the boundary prescription stated above, the globally integrated response
is then
\begin{equation}
 \Fwh_{\rm glob}(x)
 =
 \frac12
 \left[
 \frac{\pi^2}{12}
 -
 \int_0^1 C(x,\mu)\,\dd\mu
 \right].
 \label{eq:Fglobal}
\end{equation}
The first term in Eq.~\eqref{eq:Fglobal} is the globally integrated
coincident-mouth contribution, whereas $C$ collects the cross-mouth part
(defined as the global $Z$ integral of $S_{+-}$ above).
Its piecewise form records the change in the allowed mouth placements at
$q=1$.

Equations~\eqref{eq:Tpq}-\eqref{eq:Fglobal}
define the fixed-separation kernel of the matched spectator model. A treatment that
retains only the translational sector produces apparent logarithmic
image structures at special separations $r=2jL$, with
$j\in\mathbb N$. They are absent after the
exterior and reflected sectors required by Eq.~\eqref{eq:Gintdef} are
included.  The resulting kernel is continuous and numerically regular over
the displayed domain.  One-sided numerical derivatives of the two branches
agree under refinement, but no closed $C^1$ proof is claimed; pressure
derivatives are therefore checked by two routes in
Appendix~\ref{app:numerics}.

\subsection{Asymptotic behavior and numerical kernel}
\label{sec:kernel_limits}

The asymptotic behaviors of $\Fwh_{\rm glob}$ provide independent checks of
the global construction. At short separation, for example, 
\begin{equation}
 \Fwh_{\rm glob}(x)
 =
 -\frac{\pi^4}{4320}x^2
 +O(x^3),
 \qquad x\ll1,
 \label{eq:smallx}
\end{equation}
where the absence of a constant or linear term is consistent with the
original bilocal difference
$[\phi(R_+)-\phi(R_-)]^2$, which vanishes when the two mouths
coincide. In particular, $\Fwh_{\rm glob}(0)=0$.

At large separation, in turn, one obtains
\begin{equation}
 \Fwh_{\rm glob}(x)
 =
 \frac{\pi^2}{24}
 -
 \left(
 \frac{\pi^2}{16}+\frac14
 \right)\frac1x
 +O(x^{-2}),
 \qquad x\gg1,
 \label{eq:largeXF}
\end{equation}
and hence
\begin{equation}
 \lim_{x\to\infty}
 \Fwh_{\rm glob}(x)
 =
 \frac{\pi^2}{24}.
 \label{eq:Fplateau}
\end{equation}
Thus the finite large-separation plateau survives the complete global
integration, but the approach to it is smooth rather than displaying
the image singularities of an incomplete spatial-sector treatment.

\begin{figure}[t]
 \centering
 \includegraphics[width=0.6\linewidth]
 {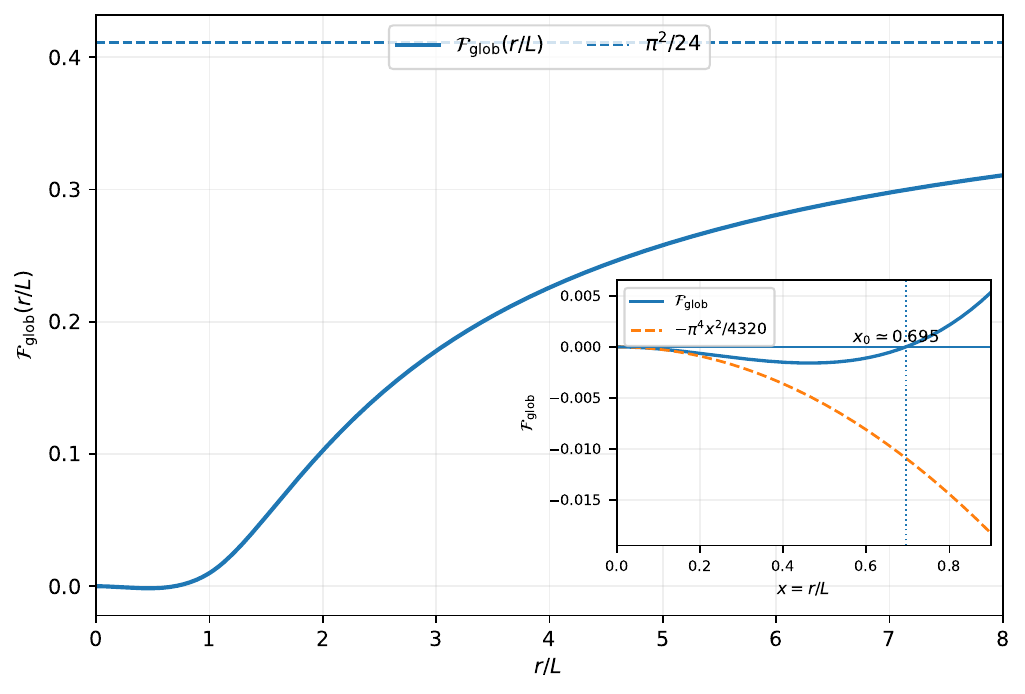}
 \caption{Matched spectator-model fixed-separation response
 $\mathcal F_{\rm glob}(x)$ versus the mouth-separation ratio $x=r/L$.
 The solid curve is the numerical kernel, the horizontal dashed line in the
 main panel marks its large-$x$ limit $\pi^2/24$, and the vertical dotted line
 marks $x_0\simeq0.695$.  The inset magnifies the short-separation region and
 compares the numerical kernel (solid) with the leading expansion
 $-\pi^4x^2/4320$ (dashed).}
\label{fig:global_kernel}
\end{figure}

Figure~\ref{fig:global_kernel} shows a fixed-pair response that is
slightly negative over a subinterval below the plate spacing and crosses
zero at $x_0\simeq0.695$ in the numerical
evaluation.  Near the origin its departure from zero is quadratic, as
required by Eq.~\eqref{eq:smallx}. At larger $x$ the curve is positive and
moves toward the analytic limit in Eq.~\eqref{eq:Fplateau}.  The zero and the
plotted trend belong to the matched point-throat model defined above.  They
do not establish a corresponding statement for finite throats or for a
different boundary matching.

\section{Distribution of spatial mouth separations}
\label{sec:distribution}

The equal-time model leaves the mouth separation as a statistical
input.  We therefore average the fixed-pair response over a normalized
radial density and compare three phenomenological families whose first two
relative moments are matched.

\subsection{Ensemble average and scale families}
\label{sec:ensemble_average}

Let $P(r)$ be normalized as in Eq.~\eqref{eq:Pnorm}.  We define the
dimensionless ensemble response $\Qwh_{\rm glob}(L)$ by averaging the
fixed-separation result in Eq.~\eqref{eq:mainfixed}:
\begin{equation}
 \frac{\overline{\Delta\EC}}{A}
 =
 \frac{na^2}{L}\,
 \Qwh_{\rm glob}(L),
 \label{eq:Eavg}
\end{equation}
where
\begin{equation}
 \Qwh_{\rm glob}(L)
 =
 \int_0^\infty
 \dd r\,
 P(r)\,
 \Fwh_{\rm glob}\!\left(\frac{r}{L}\right).
 \label{eq:Qgeneral}
\end{equation}
Equation~\eqref{eq:Qgeneral} is the central ensemble result. Since the
fixed-separation kernel is continuous and numerically regular, averaging is not required
as a regularization prescription. Its role is physical: $P(r)$
specifies the statistical distribution of the spatial nonlocality
scale sampled by the cavity. It is also important to highlight that the leading scalar effective action does not determine the microscopic
form of $P(r)$. In the present work it is therefore treated
phenomenologically, subject to normalization and to the point-throat limitation stated in
Sec.~\ref{sec:effective}. The three benchmark distributions
are used solely to test shape dependence.

Let $r_0\equiv\langle r\rangle>0$ denote the mean mouth separation, and $s$ the family identity together with any dimensionless parameters that fix its shape. Introducing the positive dimensionless variables
\begin{equation}
 u\equiv\frac{r}{r_0}>0,
 \qquad
 y\equiv\frac{r_0}{L}>0,
\label{eq:uydefs}
\end{equation}
the corresponding scale family is
\begin{equation}
 P(r;r_0,s)
 =
 \frac{1}{r_0}\,f_s(u).
\label{eq:scalefamilyP}
\end{equation}
Here, the scaled density satisfies $f_s(u)\geq0$, has unit mass and unit mean, and
is held fixed when $r_0$ is varied. However, values at $y=0$ below denote a limiting kernel value, not a distribution with zero mean separation.

Eq.~\eqref{eq:Qgeneral} then becomes
\begin{equation}
 \Qwh_{\rm glob}(L)
 =
 Q_s(y)
 =
 \int_0^\infty
 \dd u\,
 f_s(u)\,
 \Fwh_{\rm glob}(yu).
 \label{eq:Qcscale}
\end{equation}
This representation makes explicit that, for a fixed distribution
shape, the cavity response depends on the mean separation only through
the dimensionless ratio $y=r_0/L$.

\subsection{Benchmark separation distributions}
\label{sec:benchmark_distributions}

To isolate sensitivity to the shape of $P(r)$, we compare three positive-support families at the same mean $r_0$ and the same coefficient of variation (CV),
\begin{equation}
 \mathrm{CV}
 \equiv
 \frac{\sqrt{\langle r^2\rangle-r_0^2}}{r_0}.
 \label{eq:CV}
\end{equation}
Thus $\mathrm{CV}\geq0$ is dimensionless.  The continuous families
considered here use $\mathrm{CV}>0$; the zero-width case is a distributional limit.
The benchmark value is a chosen ensemble width, not a fitted measurement.
Matching these first two relative moments ensures that differences
between the resulting Casimir curves predominantly reflect higher
moments and distribution shape.

As a minimal unimodal benchmark, we use a Gaussian truncated to
positive separations. Let $\mu_G$ be a length-valued location parameter
that may have either sign, $\sigma_G>0$ be a length-valued width, and 
\begin{equation}
 \mathcal N_G
 =
 \sigma_G\sqrt{\frac{\pi}{2}}
 \left[
 1+
 \operatorname{erf}
 \left(
 \frac{\mu_G}{\sqrt2\,\sigma_G}
 \right)
 \right]>0
 \label{eq:NG}
\end{equation}
denote the length-valued normalization with $\operatorname{erf}$ meaning the error function. The normalized density is then
\begin{equation}
 P_G(r;\mu_G,\sigma_G)
 =
 \frac{1}{\mathcal N_G}
 \exp\left[
 -\frac{(r-\mu_G)^2}{2\sigma_G^2}
 \right],
 \qquad
 r>0.
 \label{eq:PG}
\end{equation}
In general $\mu_G\neq r_0$. Now, let $\varphi$ and $\Phi$ denote the standard-normal density and
cumulative distribution, respectively, and let us define the dimensionless auxiliaries
\begin{equation}
 \alpha_G
 \equiv
 -\frac{\mu_G}{\sigma_G},
 \qquad
 \lambda_G
 \equiv
 \frac{\varphi(\alpha_G)}
 {1-\Phi(\alpha_G)}.
 \label{eq:truncaux}
\end{equation}
The mean and variance of the truncated Gaussian are
written using the physical variance
$\sigma_r^2\equiv\langle r^2\rangle-r_0^2$ as
\begin{align}
 r_0
 &=
 \mu_G+\sigma_G\lambda_G,
 \label{eq:truncmean}
 \\
 \sigma_r^2
 &=
 \sigma_G^2
 \left[
 1+\alpha_G\lambda_G-\lambda_G^2
 \right].
 \label{eq:truncvar}
\end{align}
The benchmark values of $\mu_G$ and $\sigma_G$ are therefore obtained
numerically from Eqs.~\eqref{eq:truncmean} and \eqref{eq:truncvar}, with
$\sigma_r=\mathrm{CV}\,r_0$.

Let $r_\ast>0$ be an arbitrary fixed reference length; it makes the
logarithm dimensionless and is not a fitted scale.  Let
$\mu_{\ln}\in\mathbb R$ and $\sigma_{\ln r}>0$ denote the dimensionless
location and width in log space.
As a second positive-support family we use the log-normal distribution
\begin{equation}
 P_{\rm LN}(r)
 =
 \frac{1}{r\sigma_{\ln r}\sqrt{2\pi}}
 \exp\left[
 -\frac{(\ln(r/r_\ast)-\mu_{\ln})^2}
 {2\sigma_{\ln r}^2}
 \right],
 \qquad
 r>0.
 \label{eq:PLN}
\end{equation}
Matching its arithmetic mean and coefficient of variation to $r_0$
and $\mathrm{CV}$ gives
\begin{equation}
 \sigma_{\ln r}^2
 =
 \ln\!\left(1+\mathrm{CV}^2\right),
 \qquad
 \mu_{\ln}
 =
 \ln(r_0/r_\ast)-\frac12\sigma_{\ln r}^2.
 \label{eq:LNparams}
\end{equation}
Substituting Eq.~\eqref{eq:LNparams} into Eq.~\eqref{eq:PLN} cancels
$r_\ast$ from the probability density; it is only a dimensional reference,
not an additional model parameter.

For the Gamma family, let $k_\Gamma>0$ and $\theta_\Gamma>0$ denote
the dimensionless shape and length-valued scale parameters, respectively, and let
$\Gamma(\cdot)$ denote Euler's Gamma function.
The third benchmark is the density
\begin{equation}
 P_\Gamma(r)
 =
 \frac{r^{k_\Gamma-1}e^{-r/\theta_\Gamma}}
 {\Gamma(k_\Gamma)\theta_\Gamma^{k_\Gamma}},
 \qquad
 r>0.
\label{eq:PGamma}
\end{equation}
Its mean and coefficient of variation are
\begin{equation}
 \langle r\rangle=k_\Gamma\theta_\Gamma,
 \qquad
 \mathrm{CV}=\frac{1}{\sqrt{k_\Gamma}}.
\end{equation}
Hence the matched parameters are
\begin{equation}
 k_\Gamma=\frac{1}{\mathrm{CV}^2},
 \qquad
 \theta_\Gamma=r_0\,\mathrm{CV}^2.
 \label{eq:Gammaparams}
\end{equation}

The three benchmark families therefore share the same $r_0$ and
$\mathrm{CV}$
but differ in their higher moments and tail structure. This provides
a direct test of whether the Casimir response is sensitive to the
detailed form of the spatial mouth-separation distribution.

\subsection{Controlled distribution limits and total Casimir energy}
\label{sec:universalP}

For a fixed scale family $f_s(u)$ with unit mass and finite third moment,
assume also that the remainder after the quadratic term is bounded by a
constant times $x^3$ on $x\geq0$.  The short-distance expansion may then be
averaged term by term as $y=r_0/L\to0$.  Applied to
Eq.~\eqref{eq:smallx}, this gives
\begin{align}
 \Qwh_{\rm glob}(L)
 &=
 -\frac{\pi^4}{4320L^2}
 \int_0^\infty
 r^2P(r)\,\dd r
 +
 O\!\left(
 \frac{\langle r^3\rangle}{L^3}
 \right)
 \nonumber\\
 &=
 -\frac{\pi^4}{4320}
 \frac{\langle r^2\rangle}{L^2}
 +
 O\!\left(
 \frac{\langle r^3\rangle}{L^3}
 \right).
 \label{eq:universalSmallP}
\end{align}
Consequently,
\begin{equation}
 \frac{\overline{\Delta\EC}}{A}
 \simeq
 -\frac{\pi^4}{4320}
 \frac{na^2\langle r^2\rangle}{L^3},
 \qquad
 y\ll1.
 \label{eq:universalSmallEnergy}
\end{equation}
The leading short-separation response is therefore controlled by the second
moment.  The stated remainder requires the third moment; it is valid for all
three benchmark families but need not exist for a heavy-tailed distribution.

At the opposite extreme, let $f_s(u)$ be nonnegative, normalized, fixed as
$y$ varies, and free of an atom at $u=0$.
If $\Fwh_{\rm glob}$ is bounded on $x\geq0$, its pointwise limit
$\pi^2/24$ for every $u>0$ permits dominated convergence in
Eq.~\eqref{eq:Qcscale}.  The closed endpoint limits and numerical evaluation
support boundedness for the retained kernel, but a finite scan alone does not
prove it.  Under this explicit condition,
\begin{equation}
 \Qwh_{\rm glob}(L)
 \longrightarrow
 \frac{\pi^2}{24}.
 \label{eq:largeXQ}
\end{equation}
Thus
\begin{equation}
 \frac{\overline{\Delta\EC}}{A}
 \longrightarrow
 \frac{\pi^2}{24}
 \frac{na^2}{L},
 \qquad
 y\gg1.
 \label{eq:largeXE}
\end{equation}
The two limits suppress shape dependence for different reasons: the leading
small-$y$ term depends on a moment, whereas the large-$y$ limit follows from
normalization and the kernel plateau.  Neither statement is uniform over all
probability laws.  For the three fixed benchmark families, the largest
visible differences are therefore expected in the crossover region
$r_0/L=O(1)$.

Define the dimensionless nonnegative amplitude
$\eta\equiv na^2L^2\geq0$.  It is a coordinate of the ensemble scan, not
an independently fitted observable.  Combining Eqs.~\eqref{eq:Eavg} and
\eqref{eq:ordinaryCasimir} then gives
\begin{equation}
 \frac{\EC}{|\EC^{(0)}|}
 =
 -1+
 \frac{1440}{\pi^2}\,
 \eta\,
 Q_s(y).
 \label{eq:Enormalized}
\end{equation}
For a fixed-shape scale family, $Q_s(y)$ is defined by
Eq.~\eqref{eq:Qcscale}. The no-foam limit is
$\EC/|\EC^{(0)}|=-1$.

\begin{figure}[t]
 \centering
 \includegraphics[width=0.6\columnwidth]
 {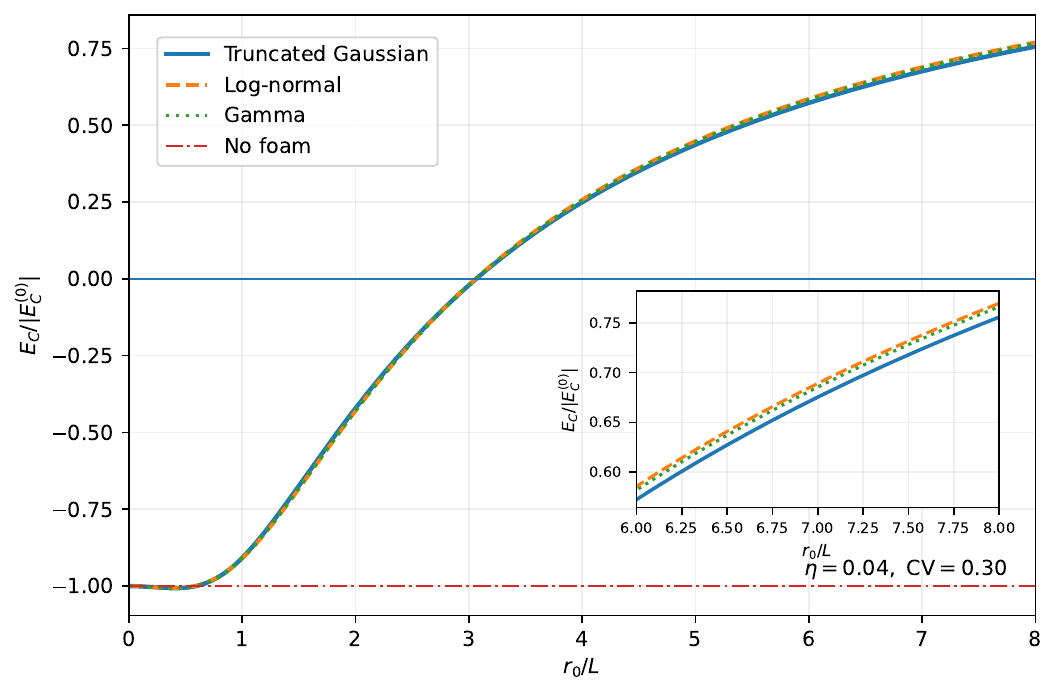}
 \caption{Total Casimir interaction energy normalized by the magnitude of the
 ordinary massless-scalar Dirichlet energy, $\EC/|\EC^{(0)}|$, versus
 $y=r_0/L$.  The curves correspond to truncated-Gaussian, log-normal, and
 Gamma mouth-separation distributions matched at
 $r_0=\langle r\rangle$ and $\mathrm{CV}=0.30$, with
 $\eta=na^2L^2=0.04$.  The dash-dotted horizontal line denotes the no-foam
 value $-1$, and the solid horizontal line denotes zero total interaction
 energy.  The inset displays $6\leq y\leq8$ on an expanded vertical scale to
 resolve the separation among the three curves.}
 \label{fig:Edistributions}
\end{figure}

Figure~\ref{fig:Edistributions} compares the three benchmark families at
$\eta=0.04$ and $\mathrm{CV}=0.30$.  Matching the mean and coefficient of
variation brings the three energy
curves close together, but the inset retains a maximum pairwise separation
of about $1.4\times10^{-2}$ in the normalized energy.  The remaining spread
is therefore small on the scale of the main panel, not zero; it records the
effect of higher moments and tail shape.  As $r_0/L$ increases, the positive
part of the fixed-pair kernel offsets the ordinary negative Casimir energy,
and the three curves cross zero near $r_0/L\simeq3.06$.  This is an energy
zero, not a force reversal.  Moreover, the plot holds $\eta$ fixed and hence
compares different ensemble parameters.

\section{Pressure, perturbative control, and regime structure}
\label{sec:pressure}

The force follows from the dimensional energy, not from a scan at fixed
$\eta$.  We therefore differentiate at fixed physical ensemble parameters
before examining the leading-order pressure and its sign.

\subsection{Pressure and asymptotic behavior}
\label{sec:pressure_derivation}

At fixed $n\geq0$, $a>0$, $r_0>0$, and distribution label $s$, the
total interaction energy per unit area is
\begin{equation}
 \frac{\EC}{A}
 =
 -\frac{\pi^2}{1440L^3}
 +
 \frac{na^2}{L}\,Q_s(y),
 \label{eq:Etotal}
\end{equation}
where $y$ and $Q_s(y)$ are defined in Eqs.~\eqref{eq:uydefs} and
\eqref{eq:Qcscale}. The physical
pressure is
\begin{equation}
 \PC
 =
 -\frac{\partial}{\partial L}
 \left(\frac{\EC}{A}\right),
 \label{eq:Pdefinition}
\end{equation}
with $n$, $a$, $r_0$, and the distribution label $s$ held fixed.  We reserve
$P(r)$ for the radial probability density and $\PC$ for the Casimir pressure.

Since $\dd y/\dd L=-y/L$, differentiation of the foam contribution gives
\begin{equation}
 \Delta\PC
 =
 \frac{na^2}{L^2}
 \left[
 Q_s(y)+yQ_s'(y)
 \right].
 \label{eq:Pgauss}
\end{equation}
We therefore define
\begin{equation}
\mathcal G_s(y)
 \equiv
 Q_s(y)+yQ_s'(y).
\label{eq:Gglobal}
\end{equation}
With this definition, $\Delta\PC=(na^2/L^2)\,\mathcal G_s(y)$, and
Eq.~\eqref{eq:Pgauss} is the compact pressure correction used below. It is useful to define the dimensionless fixed-pair pressure kernel:
\begin{equation}
 \mathcal H_{\rm glob}(x)
 \equiv
 \Fwh_{\rm glob}(x)
 +
 x\Fwh_{\rm glob}'(x).
 \label{eq:Hglobal}
\end{equation}
Thus, using Eq.~\eqref{eq:Qcscale}, Eq.~\eqref{eq:Gglobal} can then be written as
\begin{equation}
 \mathcal G_s(y)
 =
 \int_0^\infty
 \dd u\,
 f_s(u)\,
 \mathcal H_{\rm glob}(yu).
 \label{eq:GfromH}
\end{equation}
This representation is useful for establishing the large-$y$
behavior without introducing inverse moments of the separation
distribution.

Defining the dimensionless coefficients $F_\infty\equiv\pi^2/24$ and
$A_\infty\equiv\pi^2/16+1/4$ (the latter symbol is distinct from the plate
area $A$), Eq.~\eqref{eq:largeXF} takes the form
\begin{equation}
 \Fwh_{\rm glob}(x)
 =
 F_\infty-\frac{A_\infty}{x}+O(x^{-2}).
 \label{eq:Fasymgeneric}
\end{equation}
Substitution into Eq.~\eqref{eq:Hglobal} shows that the $1/x$ term
cancels identically,
\begin{equation}
 \mathcal H_{\rm glob}(x)
 =
 \frac{\pi^2}{24}
 +
 O(x^{-2}).
 \label{eq:Hlarge}
\end{equation}
For the following limit, a sufficient condition is that $f_s$ be
nonnegative and normalized, have no atom at $u=0$, and that
$\mathcal H_{\rm glob}$ be bounded on $x\geq0$.  The analytic endpoint
limits and the numerical kernel support this condition for the three
benchmarks, but the finite numerical scan is not itself a proof of global
boundedness.  Subject to this stated condition, dominated convergence in
Eq.~\eqref{eq:GfromH} yields
\begin{equation}
 \mathcal G_s(y)
 \longrightarrow
 \frac{\pi^2}{24},
 \qquad
 y\rightarrow\infty.
 \label{eq:largeXP}
\end{equation}
Unlike an expansion of $Q_s(y)$ term by term, this derivation does
not require the existence of the inverse moment
$\langle u^{-1}\rangle$.

The asymptotic foam correction to the pressure is therefore
\begin{equation}
 \Delta\PC
 \longrightarrow
 \frac{\pi^2}{24}
 \frac{na^2}{L^2}.
 \label{eq:pressurePlateau}
\end{equation}
Expressed in terms of $\eta$, the total pressure normalized by the
magnitude of the ordinary Dirichlet result becomes
\begin{equation}
 \frac{\PC}{|\PC^{(0)}|}
 =
 -1+
 \frac{480}{\pi^2}
 \eta\,
 \mathcal G_s(y).
 \label{eq:Pnormalized}
\end{equation}
Consequently,
\begin{equation}
 \frac{\PC}{|\PC^{(0)}|}
 \longrightarrow
 -1+20\eta,
 \qquad
 y\rightarrow\infty,
 \label{eq:PnormAsympt}
\end{equation}
and the leading-order asymptotic threshold for a change in the sign
of the total pressure is
\begin{equation}
 \eta_P^{\rm crit}
 =
 \frac1{20}
 =
 0.05.
 \label{eq:etaPcrit}
\end{equation}
Equations~\eqref{eq:PnormAsympt} and \eqref{eq:etaPcrit} describe the
large-$y$ limit of a fixed-$\eta$ ensemble scan.  They are not a physical
$L\to0$ limit: at fixed $n$, $a$, and $r_0$, one has
$\eta y^2=na^2r_0^2$, and decreasing $L$ can also violate $a/L\ll1$.

For comparison, the large-$y$ total energy follows from
Eq.~\eqref{eq:largeXE},
\begin{equation}
 \frac{\EC}{A}
 \longrightarrow
 \frac{\pi^2}{1440L^3}
 \left(
 -1+60\eta
 \right),
 \label{eq:Easympt}
\end{equation}
and changes sign at
\begin{equation}
 \eta_E^{\rm crit}
 =
 \frac1{60}.
 \label{eq:etaEcrit}
\end{equation}
The energy and pressure thresholds are therefore distinct. In
particular, a positive total interaction energy does not by itself
imply a repulsive Casimir force.

\subsection{Perturbative control and domain of validity}
\label{sec:pertcontrol}

The approximation scheme contains independent geometric, statistical, and
field-theoretic assumptions. The dimensionless geometric packing fraction is
\begin{equation}
\epsilon
 \equiv
 na^4
 ,\qquad \epsilon\ll1.
 \label{eq:epsilon}
\end{equation}
This parameter controls geometric overlap.  Statistical independence is
a separate assumption: the connected position correlator
$\omega(\xi,\xi')$ is set to zero in the spectator benchmark and is not made
small by Eq.~\eqref{eq:epsilon}.

The dimensionless amplitude controlling the relative size of the
foam contribution at the cavity scale is instead
\begin{equation}
 \eta
 =
 na^2L^2
 =
 \epsilon
 \left(\frac{L}{a}\right)^2.
 \label{eq:etarelation}
\end{equation}
Hence geometric diluteness, $\epsilon\ll1$, does not by itself imply
a negligible Casimir correction: $\eta$ can remain finite when
$a/L\ll1$.

A separate diagnostic follows from the momentum-space quadratic
insertion derived in Eq.~\eqref{eq:deltaK}.
For a real dummy argument $w$, the function
$\operatorname{sinc}w\equiv\sin w/w$ has global minimum approximately
$-0.2173$.
Therefore any normalized positive
$P(r)$ obeys
\begin{equation}
 0
 \leq
 1-\mathcal S(k)
 \lesssim
 1.2173.
\end{equation}
Comparing the resulting maximal free-space insertion with the lowest
Dirichlet normal-mode wave number, $k_z=\pi/L$, gives
\begin{equation}
 \chi_{\rm mode}
 \equiv
 \frac{\sup_{\mathbf k}|\delta K(\mathbf k)|}{\pi^2/L^2}
 \lesssim
 4.87\,\eta.
 \label{eq:modeestimate}
\end{equation}
This is not an operator-norm bound. Equation~\eqref{eq:deltaK} is
diagonal in the translationally invariant Fourier basis, whereas the
cavity problem is naturally expanded in Dirichlet modes.
Equation~\eqref{eq:modeestimate} is therefore used only as a free-space
mode-scale diagnostic for repeated quadratic insertions. It does not enter the derivation of the Casimir kernel.

For the plotted values, Eq.~\eqref{eq:modeestimate} gives
$\chi_{\rm mode}\lesssim0.195$ at $\eta=0.04$ and
$\chi_{\rm mode}\lesssim0.244$ at $\eta=0.05$.  Values below unity do not
bound the omitted boundary-conditioned terms, especially in the exterior
sectors.  Nor is an error bound available for the first omitted multipole in
the layers adjacent to a plate.  The resulting pressure reversal is
therefore a leading-order feature of the retained monopole kernel, not a
precision prediction.

Repeated insertions of the fixed quadratic action must be distinguished
from a response of the wormhole statistics.  In
Ref.~\cite{KirillovSavelova2015}, $V_1$ governs propagation for a prescribed
distribution, whereas the subsequent nonlinear terms describe topology
response.  The first of those terms is
\begin{equation}
 V_2[\phi]
 =
 -\frac{\pi^4}{2}
 \int \dd\xi\,a^4
 [\phi(R_+)-\phi(R_-)]^4\rho(\xi).
 \label{eq:V2}
\end{equation}
For a Gaussian scalar field, Wick theorem gives
\begin{equation}
 \left\langle[\phi(R_+)-\phi(R_-)]^4\right\rangle_0
 =
 3\left\langle[\phi(R_+)-\phi(R_-)]^2\right\rangle_0^2.
 \label{eq:V2Wick}
\end{equation}
If one ignores composite-operator subtraction and uses $L$ only as a
dimensional scale, the unrenormalized terms would count as
\begin{equation}
 \frac{\langle V_1\rangle_0}{AT_E}
 \sim
 \frac{na^2}{L},
 \qquad
 \frac{\langle V_2\rangle_0}{AT_E}
 \sim
 \frac{na^4}{L^3},
\end{equation}
and hence
\begin{equation}
 \left.
 \frac{\langle V_2\rangle_0}
      {\langle V_1\rangle_0}
 \right|_{\rm dimensional}
 =
 O\!\left(\frac{a^2}{L^2}\right).
 \label{eq:V2V1estimate}
\end{equation}
Neither Wick's identity nor this dimensional estimate determines the
boundary-subtracted quartic Casimir term: inclusion-exclusion of squared
correlators is not the square of the subtracted quadratic correlator.  The
spectator model holds the foam statistics fixed and therefore excludes
$V_2$.  A quartic extension would require a regulator, a composite-operator
prescription, and its own complete one- and two-plate subtraction.  No
numerical uncertainty is assigned from Eq.~\eqref{eq:V2V1estimate}. Repeated insertions of $V_1$, topology response through $V_2$, and
connected two-wormhole effects are distinct omissions.  The last are governed
by $\omega(\xi,\xi')$ and vanish only because statistical independence was
imposed.

The independent requirements of the present calculation may
therefore be summarized schematically as
\begin{equation}
 \epsilon\ll1,
 \qquad
 \frac{a}{L}\ll1,
 \label{eq:regime}
\end{equation}
together with the fixed-ensemble assumption.  Because the benchmark
densities have support arbitrarily close to $r=0$, the point-throat limit is
taken before the distribution average; Eq.~\eqref{eq:regime} is not a
uniform finite-$a$ support bound.  The free-space
$\chi_{\rm mode}$ estimate is not part of this validity gate because it does
not bound the boundary-conditioned operator or the exterior sectors.

As a dimensional illustration, $a/L=10^{-2}$ and
$\epsilon=4\times10^{-6}$ give $\eta=0.04$; at the same $a/L$,
$\eta=0.05$ corresponds to $\epsilon=5\times10^{-6}$.  These values only
show that a small packing fraction can coexist with a visible leading-order
amplitude.  They neither fix a physical wormhole population nor bound the
omitted terms.

\subsection{Pressure reversal and regime map}
\label{sec:pressure_results}

\begin{figure}[t]
 \centering
 \includegraphics[width=0.6\columnwidth]
 {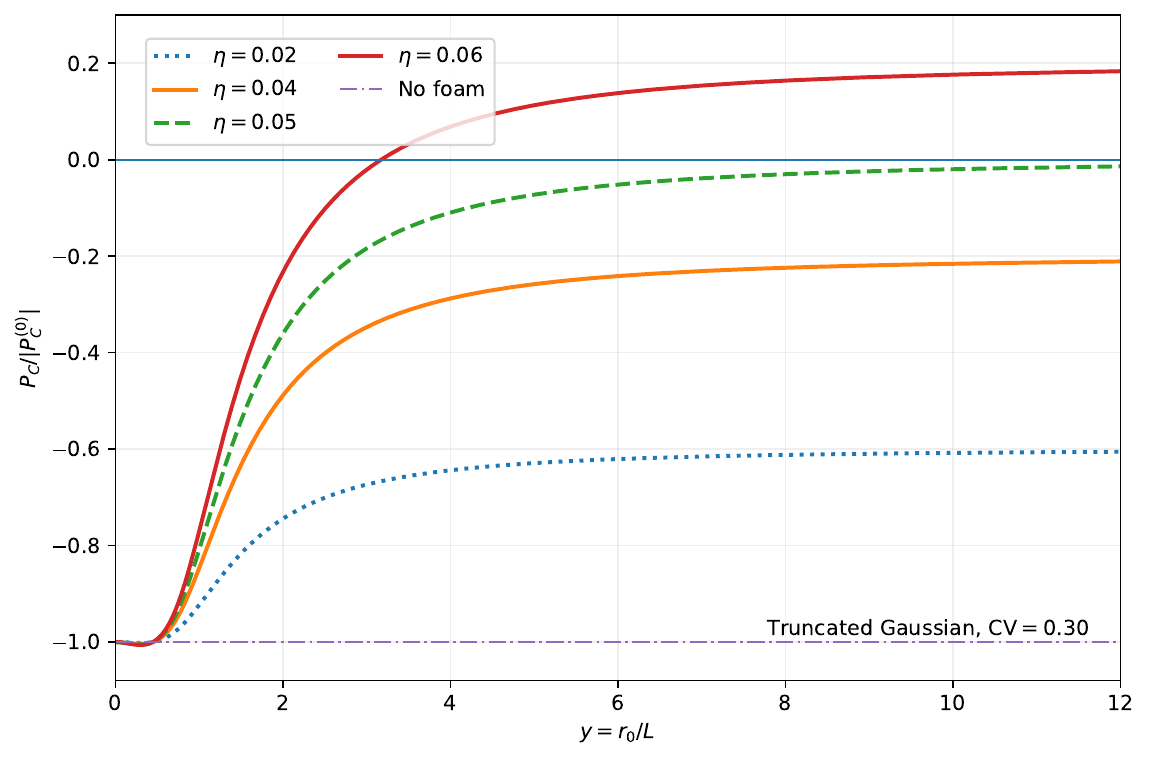}
 \caption{Total Casimir pressure normalized by the magnitude of the ordinary
 massless-scalar Dirichlet pressure, $\PC/|\PC^{(0)}|$, versus
 $y=r_0/L$ for the truncated-Gaussian distribution with
 $\mathrm{CV}=0.30$.  The fixed-$\eta$ curves use
 $\eta=0.02$, $0.04$, $0.05$, and $0.06$; the no-foam reference is
 $\PC/|\PC^{(0)}|=-1$.  The horizontal line at zero separates negative
 (attractive) from positive (repulsive) pressure.}
 \label{fig:pressure_eta}
\end{figure}

We now evaluate the pressure for the truncated-Gaussian benchmark with
$\mathrm{CV}=0.30$.  Figure~\ref{fig:pressure_eta} shows
Eq.~\eqref{eq:Pnormalized} for several values of the effective foam
amplitude. At $\eta=0.06$, the free-space diagnostic gives
$\chi_{\rm mode}\lesssim0.292$ but does not certify the
boundary-conditioned expansion.  The retained pressure vanishes at
\begin{equation}
 y_c\equiv\left(\frac{r_0}{L}\right)_c\simeq3.17.
 \label{eq:yc006}
\end{equation}
Also, across $0<y\leq12$, the
$\eta=0.02$ and $0.04$ curves remain negative and approach the
large-$y$ limits $-0.60$ and $-0.20$ of Eq.~\eqref{eq:PnormAsympt},
whereas the critical curve approaches zero without crossing it.  The
$\eta=0.06$ curve changes sign near $y\simeq3.17$ and then tends toward the
positive asymptote in Eq.~\eqref{eq:PnormAsympt}.  This is a numerical
statement about the displayed leading-order family.  It neither proves
attraction for every $y$ when $\eta<0.05$ nor controls the sign after
higher insertions, finite-throat effects, or multipoles are included.

\begin{figure}[t]
 \centering
 \includegraphics[width=0.6\columnwidth]
 {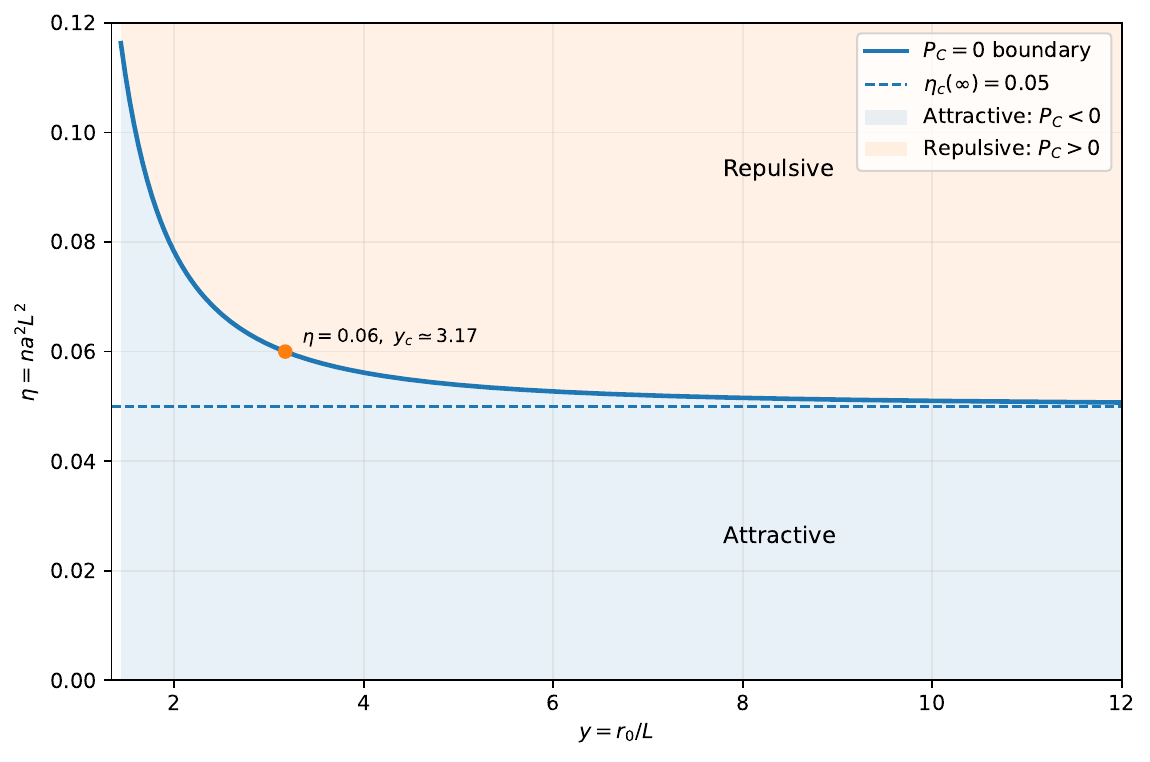}
 \caption{Leading-order pressure-sign map in the $(y,\eta)$ plane for the
 truncated-Gaussian distribution with $\mathrm{CV}=0.30$ and
 $y=r_0/L$.  The solid curve is the zero-pressure boundary defined by
 Eq.~\eqref{eq:etac_y}; the region below it has
 $\PC<0$, and the region above it has $\PC>0$.  The dashed horizontal line
 marks the large-$y$ limit $\eta_c=1/20=0.05$.  The marker at
 $(y,\eta)\simeq(3.17,0.06)$ identifies the zero of the corresponding
 pressure curve in Fig.~\ref{fig:pressure_eta}.}
 \label{fig:regime_map}
\end{figure}

Within the retained kernel, Eq.~\eqref{eq:Pnormalized} defines a
zero-pressure locus.
Whenever $\mathcal G_s(y)>0$, the boundary is
\begin{equation}
 \eta_c(y)
 =
 \frac{\pi^2}
 {480\,\mathcal G_s(y)}.
 \label{eq:etac_y}
\end{equation}
Figure~\ref{fig:regime_map} displays this curve in the $(y,\eta)$ plane:
below the boundary the retained pressure is negative, whereas above it the
pressure is positive. Since $\mathcal G_s(y)\to\pi^2/24$, the boundary approaches
$\eta_c(y)\to1/20=0.05$ at large $y$.  Thus, for the displayed benchmark,
increasing the nonlocal separation alone does not reverse the force when
$\eta<0.05$. For $\eta>0.05$, by contrast, a finite crossover can occur, and
the point $(y_c,\eta)\simeq(3.17,0.06)$ is the example displayed explicitly
in Fig.~\ref{fig:pressure_eta}.

The boundary in Fig.~\ref{fig:regime_map} classifies the sign of the
retained pressure over the plotted $(y,\eta)$ domain. It is not a dynamical
stability diagram.  Horizontal cuts reproduce the parameter scans in
Fig.~\ref{fig:pressure_eta}, but they are not trajectories generated by
moving the plates.  At fixed $n$, $a$, and $r_0$, physical variation of $L$
obeys $\eta y^2=na^2r_0^2={\rm const}$.  Such a trajectory can cross the
boundary differently from a horizontal cut.  The dashed level is only the
large-$y$ limit of the present benchmark, and the shaded signs remain
conditional on the omitted corrections listed above.

\subsection{Phenomenological interpretation}
\label{sec:phenomenology}

The physical lever in this model is the correlation scale.  Equation
\eqref{eq:etarelation} allows the leading cavity amplitude $\eta$ to remain
visible even when the packing fraction $\epsilon$ is small, while
Figs.~\ref{fig:Edistributions}-\ref{fig:regime_map} show that the response
becomes appreciable only when $P(r)$ has weight at separations comparable to
$L$.  This scale selectivity is the result supported by the scalar Dirichlet
benchmark.  It is not an experimental limit: comparison with precision
Casimir data would require an electromagnetic boundary model and a
microscopic specification of the ensemble. We also stress that the
threshold $\eta_P^{\rm crit}=0.05$ and the plateau $\pi^2/24$ belong to the
matched spectator prescription: for the confined alternative of
Appendix~\ref{app:confined} the corresponding minimum amplitudes for a pressure
reversal are $\simeq0.11$ or $\simeq0.59$, depending on how $n$ is
normalized.

\section{Conclusions}\label{sec:conclusions}

We have determined how an equal-Euclidean-time spatial bilocal interaction
changes the scalar Casimir energy and pressure between parallel Dirichlet
plates within a point-throat, leading-monopole matched spectator model.  The
foam statistics are held fixed while the scalar field obeys the boundary
conditions, and a global two-plate inclusion-exclusion subtraction over all
mouth-pair center positions isolates the finite boundary observable.

For a fixed mouth separation, the resulting dimensionless kernel is continuous
and numerically regular over the evaluated domain.  Its quadratic suppression
near coincident mouths, subsequent sign change, and finite large-separation
plateau identify the mouth separation relative to the plate spacing as the
controlling geometric scale.  Including the exterior, straddling, and reflected
sectors is essential: it removes the logarithmic image structures produced by
a restricted translation sector.

Averaging this kernel over a normalized separation density preserves the same
scale selection.  For fixed scale families satisfying the stated moment and
boundedness conditions, the short-separation response is governed at leading
order by the second moment, whereas normalization fixes the large-separation
plateau.  The three matched benchmark families yield closely spaced energy
curves, with a maximum pairwise difference of about 0.014 in normalized energy
over the resolved comparison region.  Higher moments and tail shape therefore
produce a numerically resolved residual, while the mean correlation scale
supplies the dominant response.

The pressure follows by differentiating the dimensional energy while the
physical ensemble parameters are held fixed.  This distinguishes the pressure
from a derivative along a fixed-amplitude scan and yields a leading-order
large-separation sign threshold.  In the finite truncated-Gaussian scan, the
subcritical curves remain attractive, the critical curve approaches zero, and
only the supercritical benchmark crosses into the repulsive sector.  The
result is a parameter-space classification of the retained monopole kernel,
not a physical trajectory obtained by varying the plate spacing.

Taken together, the results show that the scalar Casimir interaction can act as
a scale-selective response to spatial bilocal correlations within the declared
matched spectator model.  A microscopic separation distribution and a
first-principles boundary matching would sharpen that interpretation
(Appendix~\ref{app:confined} illustrates the size of the boundary-matching
uncertainty), while
finite-throat, higher-multipole, correlated-ensemble, and higher-insertion
calculations would quantify its accuracy.  Extending the construction to the
electromagnetic field with material response is the necessary step toward a
comparison with precision Casimir measurements.

\section*{Acknowledgments}
CRM would like to thank the Conselho Nacional de Desenvolvimento Cient\'{i}fico e Tecnol\'{o}gico (CNPq) for partial financial support, under the grant 301122/2025-3. MCA would like to thank Funda\c{c}\~{a}o Cearense de Apoio ao Desenvolvimento Cient\'{i}fico e Tecnol\'{o}gico (FUNCAP) under process No. DC3-0235-00076.01.00/24, and Conselho Nacional de Desenvolvimento Cient\'{i}fico  e Tecnol\'{o}gico - Brasil (CNPq) under process No. 304145/2025-4 for financial support. EO thanks the Funda\c{c}\~ao Cearense de Apoio ao Desenvolvimento Cient\'{i}fico e Tecnol\'ogico (FUNCAP), under the grant BP6-0241-00335.01.00/25. 

\appendix

\section{Global center integration for the bilocal kernel}
\label{app:global}

This appendix records the sector decomposition and closed expressions
used by the matched boundary prescription in Sec.~\ref{sec:cavity}.  It checks the algebra of that prescription. However, it does not derive the prescription
from the topology-averaged source theory.

For a pair of fixed separation we use the variables defined in
Eqs.~\eqref{eq:xpq} and \eqref{eq:zpm}, with $0\leq\mu\leq1$.
The parallel and normal separations are $pL$ and $qL$,
respectively. The cross-mouth part of the bilocal observable depends on the
relative location of $z_+$ and $z_-$ with respect to the two plates.
Since the foam is treated as a spectator, the center coordinate is
not restricted to the slab: the integration extends over
$-\infty<Z<\infty$. The relevant decomposition changes at $q=1$.
The ordered sectors for the two parameter ranges are listed side by side
below as complete five-row partitions.
\begin{table}[t]
\centering
\caption{Center-coordinate sectors for the two mouth positions.  The
first pair of columns applies to $q<1$, whereas the second pair applies to
$q\geq1$.  The $Z$ intervals are open and their endpoints have zero measure.
The mouth locations refer to $z_-$ and $z_+$: ``below'', ``inside'', and
``above'' mean $z<0$, $0<z<L$, and $z>L$, respectively.  At $q=1$, the
central interval in the $q\geq1$ partition is empty.}
\label{tab:center_sectors}
\begin{tabular}{cccc}
\toprule
$Z$ interval, $q<1$ & mouth locations &
$Z$ interval, $q\geq1$ & mouth locations \\
\midrule
$Z<-q/2$ & both below & $Z<-q/2$ & both below \\
$-q/2<Z<q/2$ & lower below, upper inside &
$-q/2<Z<1-q/2$ & lower below, upper inside \\
$q/2<Z<1-q/2$ & both inside &
$1-q/2<Z<q/2$ & lower below, upper above \\
$1-q/2<Z<1+q/2$ & lower inside, upper above &
$q/2<Z<1+q/2$ & lower inside, upper above \\
$1+q/2<Z$ & both above & $1+q/2<Z$ & both above \\
\bottomrule
\end{tabular}
\end{table}
Table~\ref{tab:center_sectors} displays the two complete partitions side by
side.  Within either $q$ regime, its five open intervals cover the full real
line without overlap, up to endpoints of zero measure.

Table~\ref{tab:Sintegrands} lists the integrand $S_{+-}(Z)$ of
$C(x,\mu)=\int\dd Z\,S_{+-}$ in each sector of
Table~\ref{tab:center_sectors}.  The rows follow from the component-wise prescription of
Sec.~\ref{sec:dirichlet_subtraction}: outside the slab, $G_{12}$ and the
isolated-plate functions reduce to single-image functions and vanish between
points separated by a plate, whereas $G_0$ remains the free propagator.
\begin{table}[t]
\centering
\caption{Integrand $S_{+-}(Z)$ of the cross-mouth term in each sector of
Table~\ref{tab:center_sectors}, with $g(s)\equiv1/(p^2+s^2)$ and
$z_\pm=Z\pm q/2$.  Its integral over $Z$ is the function $C(x,\mu)$.}
\label{tab:Sintegrands}
\begin{tabular}{ll}
\toprule
mouth locations ($z_-$, $z_+$) & $S_{+-}(Z)$ \\
\midrule
both inside &
$\displaystyle\sum_{m}g(q+2m)-g(q)-\sum_{m}g(2Z+2m)+g(2Z)+g(2Z-2)$ \\[3mm]
lower below; upper below or inside & $g(2Z-2)$ \\[1mm]
upper above; lower inside or above & $g(2Z)$ \\[1mm]
lower below, upper above ($q\geq1$ only) & $g(q)=1/x^2$ \\
\bottomrule
\end{tabular}
\end{table}
When both mouths are inside the slab, the first two
terms are the translational images minus the direct term, which gives
$T(p,q)$ [Eq.~\eqref{eq:app_Tsum}]. The last three are the reflections in the
two plates minus the single-plate reflections already removed by $G_{\rm int}$,
which gives $-\mathscr R_{\rm refl}(Z;p,q)$ [Eq.~\eqref{eq:app_Rintegrand}], so
the first row reproduces the sign in Eq.~\eqref{eq:Cless}.  In the other
sectors a single term survives: the reflection in the plate at $z=L$ when the
lower mouth is below the slab, the reflection in the plate at $z=0$ when the
upper mouth is above it, and the free propagator when the mouths lie on
opposite sides of the slab.  The coincident-mouth terms are treated at the end
of this appendix.

\subsection{Sector $q<1$}

When $q<1$, there exists a finite interval in which both mouths lie
between the plates. From Eq.~\eqref{eq:zpm}, the condition
$0<z_-<z_+<L$ is equivalent to
\begin{equation}
 \frac{q}{2}<Z<1-\frac{q}{2}.
 \label{eq:app_inside_interval}
\end{equation}
Its dimensionless length is therefore $1-q$. This is the origin of
the factor multiplying the translational contribution in
Eq.~\eqref{eq:Cless}.

For two points inside the slab, the translational part of the
two-plate image sum depends only on their relative displacement.
After subtracting the free-space term, the relevant dimensionless
sum is, with $m\in\mathbb Z$ the image index,
\begin{equation}
 T(p,q)
 =
 \sum_{m=-\infty}^{\infty}
 \frac{1}{p^2+(q+2m)^2}
 -
 \frac{1}{p^2+q^2}.
 \label{eq:app_Tsum}
\end{equation}
For dimensionless dummy parameters $\alpha_{\rm sum}\in\mathbb R$ and
$\beta_{\rm sum}>0$, use the standard summation identity
\begin{equation}
 \sum_{m=-\infty}^{\infty}
 \frac{1}{(m+\alpha_{\rm sum})^2+\beta_{\rm sum}^2}
 =
 \frac{\pi}{\beta_{\rm sum}}
 \frac{\sinh(2\pi\beta_{\rm sum})}
 {\cosh(2\pi\beta_{\rm sum})-\cos(2\pi\alpha_{\rm sum})},
 \label{eq:app_sumidentity}
\end{equation}

For $p>0$, setting $\alpha_{\rm sum}=q/2$ and
$\beta_{\rm sum}=p/2$ in Eq.~\eqref{eq:app_sumidentity} recovers the closed
form already defined in Eq.~\eqref{eq:Tpq}, with $x^2=p^2+q^2$ (the normal-orientation endpoint is its continuous $p\to0$ limit).
Integration over the both-inside interval
therefore produces the contribution
\begin{equation}
 (1-q)T(p,q).
 \label{eq:app_insideT}
\end{equation}

The reflected image sector depends explicitly on the center
coordinate. After the one-plate pieces have been removed by the
inclusion-exclusion prescription, the residual integrand in the
both-inside interval can be written as
\begin{equation}
 \mathscr R_{\rm refl}(Z;p,q)=\frac{\pi}{2p}\frac{\sinh(\pi p)}{\cosh(\pi p)-\cos(2\pi Z)}-\frac{1}{p^2+4Z^2}-\frac{1}{p^2+(2Z-2)^2}.
 \label{eq:app_Rintegrand}
\end{equation}
The first term contains the complete reflected two-plate image sum,
while the last two terms subtract the isolated reflections associated
with the plates at $Z=0$ and $Z=1$. Integrating $\mathscr R_{\rm refl}$ over the both-inside interval gives
the reflected residual already defined in Eq.~\eqref{eq:Rpq}. The remaining intervals contain at least one exterior mouth.  The
component-wise prescription stated in Sec.~\ref{sec:dirichlet_subtraction}
reduces their image terms to affine denominators. For real constants $b,c\in\mathbb R$, with $b\neq0$, and for $p>0$, the
required primitive is
\[
\int\frac{\dd Z}{p^2+(bZ+c)^2}
 =
\frac{1}{bp}\arctan\!\left(\frac{bZ+c}{p}\right).
\]
The prefactor $1/b$ follows from the chain rule, and direct differentiation
of the displayed primitive recovers the integrand.

The $p\to0$ endpoint is evaluated only after the
sector contributions are combined, rather than by substituting $p=0$ into
an individual primitive.
After exploiting reflection symmetry between the regions below and above
the cavity and combining the corresponding endpoint contributions, the
net exterior term becomes
\begin{equation}
 C_{\rm ext}^{(<)}(p,q)
 =
 \frac{1}{p}
 \tan^{-1}\!\left(
 \frac{p}{2-q}
 \right).
 \label{eq:app_external_less}
\end{equation} 
Explicitly, for $q<1$ the exterior sectors give
$\int_{-\infty}^{q/2}\dd Z\,g(2Z-2)+\int_{1-q/2}^{\infty}\dd Z\,g(2Z)$,
and each half-line integral equals
$(2p)^{-1}\big[\pi/2-\tan^{-1}\!\big((2-q)/p\big)\big]=(2p)^{-1}\tan^{-1}\!\big(p/(2-q)\big)$,
which sum to Eq.~\eqref{eq:app_external_less}. Combining the exterior, translational, and reflected contributions
recovers the $q<1$ branch already stated in Eq.~\eqref{eq:Cless}.

\subsection{Sector $q\geq1$}

For $q\geq1$, the interval in Eq.~\eqref{eq:app_inside_interval}
has zero or negative length, and no center position exists for which
both mouths lie simultaneously inside the cavity. The global integration
is therefore built entirely from the complementary sectors in which the
mouths occupy different regions or both lie outside the slab. Using the same
component-wise prescription in these intervals yields
\begin{equation}
 C_{>}(x,\mu)
 =
 \frac{1}{x\sqrt{1-\mu^2}}
 \tan^{-1}
 \left(
 \frac{\sqrt{1-\mu^2}}{\mu}
 \right)
 +
 \frac{\mu}{x}
 -
 \frac{1}{x^2}.
 \label{eq:app_Cgreater_arctan}
\end{equation}
For $0\leq\mu\leq1$,
\begin{equation}
 \tan^{-1}
 \left(
 \frac{\sqrt{1-\mu^2}}{\mu}
 \right)
 =
 \arccos\mu,
 \label{eq:app_trigidentity}
\end{equation}
Substitution of Eq.~\eqref{eq:app_trigidentity} into
Eq.~\eqref{eq:app_Cgreater_arctan} gives the $q\geq1$ branch in
Eq.~\eqref{eq:Cgreater}.

Equivalently, in the sectors of Table~\ref{tab:center_sectors} for $q\geq1$,
$C_>=\int_{-\infty}^{1-q/2}\dd Z\,g(2Z-2)+(q-1)/x^2+\int_{q/2}^{\infty}\dd Z\,g(2Z)
=p^{-1}\tan^{-1}(p/q)+(q-1)/x^2$. The straddling interval has length $q-1$ and
$S_{+-}=1/x^2$ there, while $p^{-1}\tan^{-1}(p/q)=\arccos\mu/(x\sqrt{1-\mu^2})$
and $(q-1)/x^2=\mu/x-1/x^2$.

\subsection{Matching at $q=1$}

At the boundary $q=1$, the domain implies $x\geq1$; one has $\mu=1/x$ and
$p=\sqrt{x^2-1}$.  The factors $(1-q)T$ and
$\mathcal R_{\rm refl}$ vanish, and both branches reduce to the common value
already displayed in Eq.~\eqref{eq:Cmatching}.  This proves continuity but
not differentiability.  As a separate numerical check, one-sided finite
differences at $x=1.1$, $2$, $3.171338$, and $8$ converge to the same branch
derivative. At a step of $10^{-5}$ the largest mismatch is below
$3\times10^{-6}$.  This test supports the numerical differentiation of the pressure
kernel but is not promoted to an analytic $C^1$ theorem.

The reflected residual in Eq.~\eqref{eq:Rpq} must be retained together
with the translational and exterior sectors.  Omitting it leaves the
restricted image structures described in Sec.~\ref{sec:fixed_kernel}. With
all adopted sectors present, the fixed-separation response is numerically
regular over the evaluated domain.

\subsection{Coincident-mouth terms and the constant $\pi^2/12$}

The terms $S_{\pm\pm}$ are given by the expressions of
Table~\ref{tab:Sintegrands} applied to a mouth and its own images, that is,
with $p=q=0$; the divergent direct term $m=0$ of the first row then cancels
against $-g(q)$.
Inside the slab the integrand reduces to
$\pi^2/12-\tfrac14\sum_{m\neq0,-1}(Z+m)^{-2}$, whose integral over $0<Z<1$ is
$\pi^2/12-\tfrac12$.  Below and above the slab the integrands are
$(2Z-2)^{-2}$ and $(2Z)^{-2}$, each integrating to $\tfrac14$.  The total is
$\pi^2/12$ per mouth, which is the constant in Eq.~\eqref{eq:Fglobal}.

\section{Numerical implementation and convergence tests}
\label{app:numerics}

The numerical evaluation applies
Eqs.~\eqref{eq:Fglobal}-\eqref{eq:GfromH} over the parameter domains shown in
the figures.  The kernel quadrature, distribution average, and pressure
calculation are implemented within the common procedure described below.

The kernel integral uses Gauss-Legendre rules with orders $N_\mu$ and
$N_Z$.  For $x>1$ the angular interval is split exactly at $\mu=1/x$;
within the $q<1$ branch, the residual integral is mapped to
$q/2\leq Z\leq1-q/2$.  Endpoint expressions are evaluated in their regular
combined limits.  The production run uses $(N_\mu,N_Z)=(100,80)$ and a
natural cubic spline through a nonuniform kernel grid: 41 points on
$0\leq x\leq0.2$, 91 on $0.2\leq x\leq2$, 101 on
$2\leq x\leq12$, and 257 on $12\leq x\leq140$, with shared endpoints
removed.  For $x<0.02$ the analytic small-$x$ expression is used only as a
stabilizing branch.  Direct evaluation of the closed integral across the
overlap $0.005\leq x\leq0.05$ verifies the expected $O(x^3)$ departure; at
$x=0.02$ the absolute difference is $2.7\times10^{-7}$.

The distribution average in Eq.~\eqref{eq:Qcscale} uses $N_u=140$
Gauss-Legendre nodes up to the $1-10^{-12}$ quantile $u_{\max}$ of each
benchmark.  The retained mass differs from unity by less than
$1.1\times10^{-12}$ in all three cases.  The
energy, pressure, and regime scans use 161, 237, and 211 uniformly spaced
values of $y$ over $0\leq y\leq8$, $0.2\leq y\leq12$, and
$1.5\leq y\leq12$, respectively.

The pressure is calculated both by differentiating a spline of $Q_s(y)$
and by the direct integral in Eq.~\eqref{eq:GfromH}.  These routes share the
same fixed-pair kernel, so their agreement tests numerical differentiation,
not the physical derivation of the kernel.  In the production run they give
$y_c=3.17133616$ and $3.17133625$ for $\eta=0.06$; their difference is
$9.4\times10^{-8}$.  Convergence of the displayed precision is summarized
below.

A higher-order run with $(N_\mu,N_Z,N_u)=(180,140,220)$ gives
$\mathcal G_s(2)=0.2630002$ and $y_c=3.171338$, consistently with the
convergence shown in Table~\ref{tab:numerical_convergence}.  At an exact root,
$\mathcal G_s(y_c)=\pi^2/(480\eta)$ by definition, so $y_c$ is the independent
root diagnostic.  The four resolutions support $y_c\simeq3.17$ at the stated
precision.  The small-$x$ and large-$x$ analytic checks, branch matching,
quadrature refinement, and two pressure routes test different numerical
failure modes; none of them establishes the source-to-boundary matching
assumption.

\begin{table}[t]
\centering
\caption{Quadrature convergence for the dimensionless kernel value
$\mathcal G_s(2)$ and for the pressure root at $\eta=0.06$.  The first
column gives the angular, center-coordinate, and distribution quadrature
orders $(N_\mu,N_Z,N_u)$; the last column reports the root obtained in each
run.  The three tabulated resolutions show convergence at the displayed
precision.}
\label{tab:numerical_convergence}
\begin{tabular}{ccc}
\toprule
$(N_\mu,N_Z,N_u)$ & $\mathcal G_s(2)$ & $y_c$ at $\eta=0.06$ \\
\midrule
$(70,55,100)$   & \;0.2630001 & 3.171340 \\
$(100,80,140)$  & \;0.2630003 & 3.171336 \\
$(130,100,180)$ & \;0.2630002 & 3.171338 \\
\bottomrule
\end{tabular}
\end{table}

\section{Alternative boundary prescription: pairs confined to the slab}
\label{app:confined}

The kernel of Sec.~\ref{sec:fixed_kernel} follows from the matched spectator
prescription, in which the pair center is integrated over the whole normal
line and the field is disconnected across each plate.  To gauge how much of
the sign structure depends on that postulate, we compare it with a second,
equally simple alternative prescription: only pairs whose two mouths both lie inside the
slab are retained, with the same $G_{\rm int}$ and the same prefactor
$\pi^2na^2$.  The $Z$ integral is then restricted to $q/2<Z<1-q/2$, which is
empty for $q\geq1$, and Eq.~\eqref{eq:mainfixed} defines a kernel
$\Fwh_{\rm conf}(x)$,
\begin{equation}
 \Fwh_{\rm conf}(x)
 =\frac14\int_0^{\min(1,1/x)}\!\!\dd\mu\int_{q/2}^{1-q/2}\!\!\dd Z\,
 \Big[\mathcal S\big(Z+\tfrac q2\big)+\mathcal S\big(Z-\tfrac q2\big)
 -2S_{+-}(Z;p,q)\Big],
 \label{eq:Fconf}
\end{equation}
where $S_{+-}$ is the ``both inside'' expression of
Table~\ref{tab:Sintegrands}, and $\mathcal S(z)$ is the coincident-mouth term
$S_{\pm\pm}$ of a mouth at height $z$ (in units of $L$) inside the slab,
$\mathcal S(z)=\pi^2/12-\tfrac14\big[\psi_1(1+z)+\psi_1(2-z)\big]$, with
$\psi_1$ the trigamma function.  Two normalizations are considered.  In the
first, $n$ and $P(r)$ are those of the main text.  In the second, $n$ counts
only pairs that fit in the slab, so the kernel is divided by the length
available to the pair center, averaged over orientations,
$\Lambda(x)=\int_0^{\min(1,1/x)}(1-x\mu)\,\dd\mu$, i.e.\ $\Lambda=1-x/2$ for
$x\leq1$ and $\Lambda=1/(2x)$ for $x>1$:
$\Fwh_{\rm cond}\equiv\Fwh_{\rm conf}/\Lambda$.

\begin{figure}[t]
 \centering
 \includegraphics[width=0.99\columnwidth]{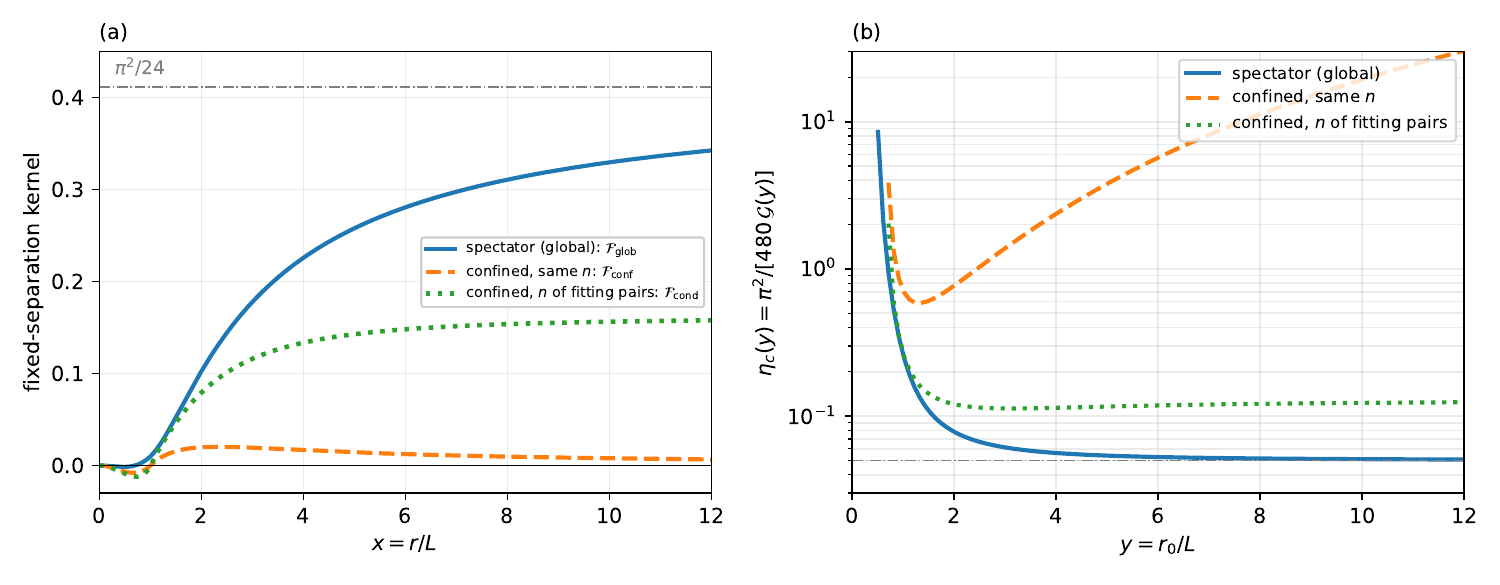}
 \caption{Spectator (global) and confined prescriptions.  (a) Fixed-separation
 kernels $\Fwh_{\rm glob}$, $\Fwh_{\rm conf}$, and
 $\Fwh_{\rm cond}=\Fwh_{\rm conf}/\Lambda$; the dash-dotted line marks
 $\pi^2/24$.  (b) Zero-pressure boundary
 $\eta_c(y)=\pi^2/[480\,\mathcal G(y)]$ for the truncated-Gaussian benchmark
 with $\mathrm{CV}=0.30$ (shown where $\mathcal G>0$); the dash-dotted line
 marks $\eta=0.05$.}
 \label{fig:confined}
\end{figure}

Numerically, the confined kernel shares the sign structure of
$\Fwh_{\rm glob}$: it starts negative, $\Fwh_{\rm conf}\simeq-0.057\,x^2$
at small $x$ (against $-\pi^4x^2/4320\simeq-0.0225\,x^2$ in the spectator
case), crosses zero near $x\simeq1.0$ (against $0.695$), reaches a maximum
$\simeq0.020$ near $x\simeq2$, and then decays as $\simeq0.08/x$, so there is
no plateau.  The conditional kernel $\Fwh_{\rm cond}$ tends to a plateau
$\simeq0.16$, to be compared with $\pi^2/24\simeq0.41$.  We checked
Eq.~\eqref{eq:Fconf} against a direct sum over images at $x=0.6$ and $1.7$
(agreement to eight digits). Averaging over the truncated-Gaussian benchmark with $\mathrm{CV}=0.30$ and
repeating the pressure analysis of Sec.~\ref{sec:pressure_results}, the
minimum amplitude for which $\PC$ can change sign is
$\eta_c^{\min}=\pi^2/[480\,\max_y\mathcal G(y)]$.  The results are collected
in Table~\ref{tab:confined} and Fig.~\ref{fig:confined}.

The qualitative structure-negative response at small separation, a sign
change near $r\sim L$, and an appreciable effect only when $P(r)$ has weight
at separations comparable to $L$ is common to the prescriptions.  The
existence of a plateau, and the numerical value of the minimum amplitude, are
not: they range from $\simeq0.05$ to $\simeq0.59$.  For the confined kernel
with the same $n$, the reversal requires $\eta\simeq0.59$, where
$\chi_{\rm mode}\lesssim2.9$ exceeds unity, so the first-order expansion is
not expected to be reliable and no controlled repulsive regime is found in
that case.  Both prescriptions are postulates, and neither is
derived from the topology-averaged source theory. The results of the main
text should therefore be read as conditional on the matched spectator
prescription.

\begin{table}[t]
\centering
\caption{Minimum foam amplitude for a pressure sign change, truncated-Gaussian
benchmark with $\mathrm{CV}=0.30$ and $0.3\leq y\leq12$, for the three
prescriptions, and the corresponding free-space diagnostic
$\chi_{\rm mode}\lesssim4.87\,\eta$ of Eq.~\eqref{eq:modeestimate}.}
\label{tab:confined}
\begin{tabular}{lccc}
\toprule
prescription & kernel at large $x$ & $\eta_c^{\min}$ & $\chi_{\rm mode}$ at $\eta_c^{\min}$ \\
\midrule
spectator (global) & plateau $\pi^2/24$ & $0.051$ ($\to0.05$) & $0.25$ \\
confined, same $n$ & $\simeq0.08/x\to0$ & $0.59$ & $2.9$ \\
confined, $n$ of fitting pairs & plateau $\simeq0.16$ & $0.11$ & $0.55$ \\
\bottomrule
\end{tabular}
\end{table}

\bibliographystyle{apsrev4-2}
\bibliography{Ref}

\end{document}